\documentclass{article}

\PassOptionsToPackage{numbers}{natbib}
\usepackage[main, final]{neurips_2026}

\usepackage[utf8]{inputenc}
\usepackage[T1]{fontenc}
\usepackage{hyperref}
\usepackage{url}
\usepackage{booktabs}
\usepackage{amsfonts}
\usepackage{nicefrac}
\usepackage{microtype}
\usepackage{xcolor}
\usepackage{amsmath}
\usepackage{amssymb}
\usepackage{mathtools}
\usepackage{amsthm}

\usepackage{algorithm}
\usepackage{algorithmic}
\usepackage{xspace}
\usepackage{graphicx}
\usepackage{wrapfig}
\usepackage{url}
\usepackage{color}
\usepackage{multirow}
\usepackage{gensymb}
\usepackage{longtable}
\usepackage{tikz}
\usepackage{dsfont}
\usepackage{framed}
\usepackage{enumitem}
\usepackage{bm}
\usepackage{pifont}
\usepackage{natbib}
\usepackage{wrapfig}
\usepackage{subcaption}
\usepackage[capitalize,noabbrev]{cleveref}

\theoremstyle{plain}
\newtheorem{theorem}{Theorem}

\newtheorem{lemma}[theorem]{Lemma}

\theoremstyle{definition}
\newtheorem{definition}{Definition}
\newtheorem{assumption}{Assumption}
\theoremstyle{remark}

\newcommand{\ouralg}{\text{PPOA}\xspace}
\newcommand{\ouralgftrl}{\text{PPOA-FTRL}\xspace}
\newcommand{\ouralgdmd}{\text{PPOA-DMD}\xspace}
\newcommand{\oadp}{\text{DPA}\xspace}
\newcommand{\greedy}{\text{Greedy}\xspace}
\newcommand{\dmd}{\text{DMD}\xspace}
\newcommand{\ftrl}{\text{FTRL}\xspace}
\newcommand{\ndmd}{\text{NDMD}\xspace}
\newcommand{\nftrl}{\text{NFTRL}\xspace}
\newcommand{\pridown}{\text{Pri-DOWM}\xspace}

\title{Online Allocation with Differential Privacy}

\author{
  Jianyi Yang\thanks{Correspondence to Jianyi Yang <jyang66@uh.edu>} \\
  University of Houston\\
  \texttt{jyang66@uh.edu} \\
  \And
  Xingyu Zhou \\
  Wayne State University \\
  \texttt{xingyu.zhou@wayne.edu} \\
  \AND
  Mostafa Mushsharat \\
  University of Houston \\
  \texttt{mmushsha@cougarnet.uh.edu} \\
  \And
  Shaolei Ren \\
  University of California,  Riverside \\
  \texttt{shaolei@ucr.edu} \\
}

\begin{document}

\maketitle

\begin{abstract}
 We study private online allocation problems, where allocation decisions must satisfy differential privacy (DP) to protect sensitive user information while optimizing performance under limited resources.  We first formalize differential privacy notions suited to different privacy requirements in online allocation and establish the fundamental performance limits of any algorithm under these definitions. Based on these insights, we propose \ouralg, a privacy-preserving meta algorithm for online allocation, and establish sufficient conditions under which \ouralg preserves Joint DP (JDP) or Local DP (LDP) while achieving asymptotically near-optimal performance. Building on these conditions, \ouralg can be instantiated into a variety of concrete algorithms. In particular, we present the specific designs, which include \ouralgdmd that preserves JDP and LDP and \ouralgftrl that preserves JDP. We analyze their performance under both adversarial and stochastic settings and characterize the fundamental trade-offs between DP and allocation performance. Finally, we demonstrate the superior performance of the proposed algorithms via numerical experiments on AI model routing in battery-powered edge systems.
\end{abstract}

\section{Introduction} 
Online allocation, with applications such as resource-constrained computing~\cite{edge_computing_energy_constraints_jiang2022joint,energy_management_battery_rakhmatov2003energy}, 
online advertising~\cite{online_matching_mehta2013online,learning_auctions_balseiro2019learning}, 
and inventory management~\cite{fair_allocation_public_goods_fain2018fair,fair_allocation_perishable_resources_banerjee2023online}, is a fundamental problem where an action $x_t\in\mathcal{R}^d$ is decided for the user at round $t\in [T]$ with a reward function $f_t:\mathcal{X}_t\rightarrow \mathcal{R}_+$, and a resource consumption function $b_t:\mathcal{X}_t\rightarrow \mathcal{R}_+^m$ where $\mathcal{R}_+$ represents the positive real domain and $m$ is the number of resources. The objective is to maximize the cumulative reward while ensuring the element-wise constraints given the resource budget $B\in \mathcal{R}_+^m$ over $T$ rounds, formulated below.
\begin{equation}\label{eqn:objective_offline}
\begin{split}
       \mathrm{OPT}(\mathbf{y}) \coloneq \max_{x_{1:T}} \sum_{t=1}^Tf_t(x_{t})\; \; \mathrm{s.t.}, \;\; \sum_{t=1}^T b_t(x_t)\leq B \;\text{ and }\; x_t\in\mathcal{X}_t,\;\;\forall t\in[T]
  \end{split}
\end{equation}

The presence of budget constraints inherently couples the decisions across different requests, so the decision for a request contains the information of other requests. 
This coupling induces a resource-constraint-based side channel through which an attacker can infer other users' contextual information by observing the actions. The privacy risks are further illustrated by application examples in Section~\ref{sec:application}. 

Differentially private resource allocation has been studied both for problems with specific privacy requirements \cite{DR_allocation_chen2023differentially,DP_allocation_blockchain_guo2021differential} and for more general settings aimed at achieving algorithmic performance guarantees \cite{noisy_DMD_chen2024noisy,JDP_packing_huang2018near}. Among the latter, \cite{noisy_DMD_chen2024noisy} propose an \textit{offline} joint differential private algorithm for \textit{convex} objectives that guarantees expected performance over the algorithm’s randomness, while \cite{JDP_packing_huang2018near} develop an online joint differential private algorithm under a \textit{random} arrival model. However, neither work addresses the adversarial online setting with worst-case competitive performance guarantees that hold uniformly over all context sequences, with failure probabilities vanishing as the time horizon grows.
This paper aims to answer the question: 

\emph{What are the fundamental limits of online allocation under different DP requirements, and how can we achieve meaningful performance guarantees in the adversarial setting under these DP constraints?}

Answering this question requires solving the following challenges.
\textbf{(1) Understanding the performance limits of privacy requirements.}
Online allocation admits multiple privacy notions under different settings, making it essential to characterize the fundamental performance limits imposed by these definitions.
\textbf{(2) Private coordination under global constraints.}
Online allocation requires inter-user coordination through shared information, such as remaining budgets or dual variables, to enforce global resource constraints and optimize overall performance. However, such coordination can leak sensitive user information, and the way it is privatized critically affects performance. Currently, there is a lack of a unified meta-framework that privatizes coordination variables while provably maintaining feasibility and achieving strong allocation performance.
\textbf{(3) Adversarial performance guarantee under DP.}
Maintaining competitive performance under privacy constraints is
particularly challenging for adversarial context sequences, as the
randomness introduced by DP mechanisms must be controlled to obtain
guarantees with vanishing failure probabilities. Directly applying private online learning methods, such as noisy Dual Mirror Descent~\cite{noisy_DMD_chen2024noisy}, within such a framework fails to provide adversarial guarantees, as unbounded privatization noise can accumulate over time and lead to large performance gaps.
We make the following primary contributions. 

 \textbullet\;  We introduce the standard DP, Joint DP, and Local DP notions into online allocation problems and characterize the fundamental tension between DP and allocation performance by establishing best-achievable performance. 
 Our analysis reveals that no algorithm can achieve
near-optimal performance guarantees under $(\epsilon,\delta)$-DP
(Theorem~\ref{thm:DP_lower_bound}), and we
characterize the DP-performance trade-offs by establishing the best-achievable competitive ratio under JDP and LDP
(Theorem~\ref{thm:JDP_CR_upper_bound}).

 \textbullet\; We propose a meta framework for online allocation with JDP or LDP (\ouralg; Algorithm~\ref{alg:algorithm_meta}) and establish sufficient conditions under which \ouralg preserves $(\epsilon,\delta)$-JDP or $(\epsilon,\delta)$-LDP (Theorem~\ref{thm:DP_meta}), satisfies the resource constraints with any prescribed high probability, and achieves near-optimal performance simultaneously in both adversarial and i.i.d. stochastic settings (Theorem~\ref{thm:reward_meta}). \ouralg can be instantiated by incorporating different DP designs that satisfy the conditions.

   \textbullet\; 
We design two instantiations of \ouralg, including \ouralgdmd (Algorithm~\ref{alg:algorithm_local}) which preserves both $(\epsilon,\delta)$-JDP and $(\epsilon,\delta)$-LDP (Theorem~\ref{thm:privacy_local}) and the \ouralgftrl (Algorithm~\ref{alg:algorithm_ftrl}) specifically for $(\epsilon,\delta)$-JDP preservation (Theorem~\ref{thm:privacy_ftrl}). They are proved to achieve asymptotic optimal performance with different DP-performance trade-offs (Theorem~\ref{thm:reward_central},Theorem~\ref{thm:reward_local}). They are empirically evaluated by sufficient ablations and baseline comparisons on the battery-powered edge computing application.

\section{Application Examples}\label{sec:application}
We summarize two applications here while leaving their details and an additional example about \emph{budget-aware tool calls in agent service}  in Section~\ref{sec:more_application}.

\textbf{Task Offloading for Battery-Powered Edge Computing}
In battery-powered edge computing systems, IoT devices handle tasks from different users under limited energy budgets and decide in real time to offload computation to cloud or other edge servers to balance latency, performance, and energy consumption \cite{edge_computing_energy_constraints_jiang2022joint,edge_survey_luo2021resource,yang2018mobile}. 
However, the shared energy constraint introduces a side channel: the decision for one request can implicitly reveal information about prior requests. This creates privacy risks: by observing the offloading decisions for its own tasks, a malicious user can infer the information of other users. Thus, privacy-preserving offloading algorithms are needed to protect user information while maintaining high performance.

\textbf{Energy Management in Multi-Tenant Data Centers}
A multi-tenant data center is a decentralized system where the tenants are co-located in one data center and supported by shared resources but decide their own computing configurations locally \cite{islam2016market,dai2015energy,chen2015greening}.  During peak hours, the multi-tenant data center must reply on a shared energy storage system for computation, which determines a shared energy budget for all the tenants. During these hours, each tenant determines its configurations based on its workload characteristics while adhering to a shared global energy constraint. Enforcing this constraint requires servers to exchange coordination information, which can create side channels that leak sensitive information about other tenants’ workloads. To mitigate such risks, tenants must privatize their shared coordination information, motivating the local DP problem in resource allocation.

\section{Related Work}\label{sec:related works}

\paragraph{Online Allocation} 
Online allocation has been widely studied. Some works consider the stochastic settings where the inputs are drawn from a distribution  \cite{near_optimal_online_allocation_atochastic_devanur2019near,fair_allocation_nonstationary_liao2022nonstationary,online_allocation_unknown_arrival_an2024best, fair_online_allocation_balseiro2021regularized}. Also, online algorithms have been proposed to guarantee the worst-case performance given adversarial inputs \cite{dual_mirror_descent_online_allocation_balseiro2020dual,knapsack_evcharging_sun2020competitive,knapsack_sun2023online,CR_pursuit_multi_constraint_lin2022competitive}.  Recently, algorithms are designed to achieve the best performances in both adversarial and stochastic settings \cite{dual_mirror_descent_online_allocation_balseiro2020dual,knapsack_post_reveal_castiglioni2022online}.  The privacy risks of online allocation are often overlooked by these methods while we design algorithms to achieve near-optimal performance even under DP guarantees. 

\paragraph{Differential Private Online Algorithms}
 Differential private algorithms have been studied for many online problems such as online learning \cite{DP_online_learning_price_agarwal2017price,DP_FTRL_kairouz2021practical,DP_FTRL_Federated_learning_xu2023federated},  online bandit learning \cite{DP_MAB_tossou2016algorithms,privay_partial_info_azize2022privacy,DP_MAB_cost_basu2019differential}, and reinforcement learning \cite{vietri2020private, LDP_RL_garcelon2021local,chowdhury2022differentially,JDP_adversarial_MDP_baiprivate, DP_RL_linear_zhou2022differentially, DR_RL_rio2024differentially}.  
Private online algorithms are also studied under different DP definitions. Some works consider a relaxed JDP definition which provides strong enough privacy guarantee for their settings \cite{JDP_bandits_shariff2018differentially,JDP_packing_huang2018near,JDP_adversarial_MDP_baiprivate}. Moreover, LDP mechanisms \cite{LDP_bandits_zheng2020locally,ren2020multi,LDP_DP_MDP_liao2023locally,LDP_RL_garcelon2021local,LDP_robust_bandits_zhou2024locally} are proposed to privatize information shared by each local user. The study of DP for online allocation remains nascent and presents great algorithmic challenges for performance guarantee with budget constraint satisfaction.

\textbf{Private Resource Allocation}. 
Private resource allocation problems have been studied in many settings.  \cite{DR_allocation_chen2023differentially} reveal the privacy risk in an offline  resource-constrained networked system and provides several privatizing mechanisms.
\cite{noisy_DMD_chen2024noisy} develop a noisy dual mirror descent algorithm for JDP preserving offline resource allocation with convex objectives, which achieves a performance bound of $\widetilde{\mathcal{O}}({\frac{1}{\epsilon}})$ with bounded constraint violations.
\cite{JDP_packing_huang2018near}  provide a jointly differentially private multiplicative weight algorithm for a packing problem with linear objective, achieving a bound of $\widetilde{\mathcal{O}}({\frac{1}{\epsilon}})$. Although an online implementation is provided by \cite{JDP_packing_huang2018near}, it requires the assumption of random arrival. While \cite{DP_allocation_blockchain_guo2021differential} studies online allocation problem, they focus on protecting against two specific inference attacks in an auction setting, and its mechanism only privatizes a threshold parameter arising from its particular algorithm design, which does not protect the underlying user context information.

\section{Performance Metrics, Privacy Definitions and Performance Limits}\label{sec:formulation}
This section introduces the privacy definitions and performance metrics for online allocation. We first introduce some notations and assumptions for the online allocation problem in \eqref{eqn:objective_offline}. 
We denote the context for user at round $t$ as $y_t=\{\mathcal{X}_t, f_t,b_t\}$ and $\bm{y}=[y_1,y_2\cdots, y_T]$ which satisfy Assumption~\ref{asm:reward_cost_assumptions}. 
\begin{assumption}\label{asm:reward_cost_assumptions}
We assume that the reward function and resource consumption at each round are upper bounded, i.e. $\forall t\in [T], 0\leq f_t(x_t)\leq \bar{f}$, and $\forall t\in [T],\forall j\in [m]$, $0\leq [b_t(x)]_j\leq \bar{b}_j$ and $\sup \|b_t(x)\|_2\leq \bar{b}$. Also for each $j\in[m]$ the consumption upper bound $\bar{b}_j$ is no smaller than the time-averaged budget $\rho_j=B_j/T$.  The action set $\mathcal{X}_t$ always includes $0$ to represent a null action, with $b_t(0)=0$ and $f_t(0)=0$ accordingly. We \textit{do not} assume other function assumptions like continuity, convexity or smoothness.
\end{assumption}

\subsection{Performance Metrics}
To guarantee DP, we need to introduce random noise in algorithm design, which makes it difficult to strictly guarantee constraint feasibility given any finite horizon.  We therefore require that, the algorithms can be configured to satisfy the budget constraints with any prescribed failure probability as is defined below. 
\begin{definition} [Constraint feasibility w.h.p.]\label{def:feasibility}
An online allocation algorithm $\mathcal A$ satisfies the constraint feasibility with high probability (w.h.p.) if for any prescribed failure probability $\zeta\in (0,1)$ and any fixed
context sequence $\bm y\in\mathcal Y^T$, with probability at least $1-\zeta$ over the randomness of algorithm $\mathcal A$, the actions $\{x_1,\cdots, x_T\}$ of algorithm $\mathcal A$  satisfy the constraint $\sum_{t=1}^Tb_t(x_t)\leq B$. 
\end{definition}

The performance metrics for online allocation algorithms under both stochastic and adversarial settings are defined for algorithms that satisfy the constraint feasibility w.h.p. (Definition~\ref{def:feasibility}). 
In the stochastic setting, the user contexts $\bm{y}$ are drawn from a distribution $\mathcal{P}_T$.  Under this setting, the objective is to minimize the expected regret expressed as 
\(
\mathbb{E}\left[\mathrm{OPT}(\bm{y})- F_T^{\mathcal{A}}(\bm{y}) \right]
\) where $F_T^{\mathcal{A}}(\bm{y})=\sum_{t=1}^T f_t(x_t)$ is the total reward given an algorithm $\mathcal{A}$. We mainly express the theoretical results for the stationary stochastic setting where the context of each user $y_t$ is drawn independently from the same distribution $\mathcal{P}$, but the results can be generalized to non-stationary setting without much effort.

More importantly, we focus on \textbf{worst-case performance in the adversarial setting}, seeking guarantees that hold uniformly over all context sequences, with failure probabilities over the algorithm's randomness vanishing as the time horizon grows. This formulation accommodates arbitrary context sequences while accounting for the stochastic behavior introduced by privacy
mechanisms. Such performance guarantees are particularly relevant to safety-critical applications \cite{noor2020survey,huang2023safe,ren2020resource}.
In this setting, worst-case performance is typically evaluated using the 
competitive ratio 
\cite{online_matching_mehta2013online,eOBM_fahrbach2022edge,knapsack_evcharging_sun2020competitive,knapsack_sun2023online}, 
defined as the minimum ratio between the algorithm reward
and the optimal offline reward over context sequences and the algorithm randomness, i.e.,
\(
\inf 
\frac{F_T^{\mathcal{A}}(\bm{y})}{\mathrm{OPT}(\bm{y})}.
\)
However, guaranteeing the competitive ratio becomes extremely challenging 
due to the randomness introduced by DP. 
Thus, we relax the competitive ratio to  
\textit{asymptotic competitive ratio in probability} defined below to illustrate meaningful guarantee.

\begin{definition}[Asymptotic competitive ratio in probability]
\label{def:cr}
A random algorithm $\mathcal A$ has an asymptotic competitive ratio of at least $1/\alpha$ in probability, for a fixed $\alpha\geq 1$, if, for every $\kappa>0$,
\[
\lim_{T\to\infty}
\sup_{\bm y\in\mathcal Y^T}
\Pr_{\mathcal A}
\left[
\frac{
\mathrm{OPT}(\bm y)
-\alpha F_T^{\mathcal A}(\bm y)
}{T}
>\kappa
\right]
=0,
\]
where the probability is over the algorithm's internal randomness.
\end{definition}
 The definition of asymptotic competitive ratio in probability adapts the notion of convergence in probability~\cite{convergence_in_probability_pishro2014introduction} to competitive performance. An algorithm with a positive asymptotic competitive ratio in probability achieves the corresponding competitive guarantee up to a sublinear additive error uniformly over all input sequences, with a failure probability that vanishes as the time horizon grows.

\subsection{Privacy Definitions and Performance Limits}\label{sec:DP_metrics}

The actions of online allocation \eqref{eqn:objective_offline}, if observed by malicious attackers, may disclose sensitive contextual information about rewards and resource consumption, thereby introducing potential privacy risks. The privacy requirements are formally specified by the following privacy notions.
\paragraph{Differential Privacy}

The standard $(\epsilon, \delta)$-DP defined below is commonly adopted.
\begin{definition}[$(\epsilon, \delta)-$DP]\label{def:DP}
 For any $\epsilon\geq 0$, $\delta\in(0,1)$, an algorithm $\mathcal{A}: \mathcal{Y}^T\rightarrow \mathcal{X}^T$, where $\mathcal{Y}^T=\mathcal{Y}_1\times \cdots\times \mathcal{Y}_T$ and $\mathcal{X}^T=\mathcal{X}_1\times\cdots\times \mathcal{X}_T$ are compositions of $T$-round inputs and outputs,  is ($\epsilon, \delta$)-differential private if for any pair of neighboring sequence $\bm{y}, \bm{y}'\in \mathcal{Y}^T$ that differ in the input of one user $t$, and any subset of $O\subset \mathcal{X}^T$, $\Pr(\mathcal{A}(\bm{y})\in O)\leq e^{\epsilon}\cdot \Pr(\mathcal{A}(\bm{y}')\in O)+\delta$.
\end{definition}
DP with a smaller $\epsilon$ requires the outputs of two neighboring sequences are less distinguishable, thereby providing a stronger privacy guarantee. In addition, allowing $\delta \in (0,1)$ relaxes this requirement by permitting the privacy guarantee to fail with probability at most $\delta$.
However, $(\epsilon, \delta)$-DP significantly restricts the variation of actions in response to user contexts, making it impossible to achieve near-optimal reward performance. This is substantiated by the following Theorem. 
\begin{theorem}\label{thm:DP_lower_bound}
For any $\epsilon>0, 0<\delta<1$, \textbf{no} online allocation algorithm that preserves $(\epsilon, \delta)$-DP can achieve a sublinear regret. 
In addition, the asymptotic competitive ratio in probability (Definition~\ref{def:cr}) of an $(\epsilon,\delta)$-DP algorithm is at most
\(\delta.\)
\end{theorem}
The proof of Theorem \ref{thm:DP_lower_bound} are deferred to Appendix \ref{sec:proofs_lower_bound}. 
Intuitively, the standard $(\epsilon,\delta)$-DP restricts the online allocation algorithm to adapt its actions in response to context variations, which incur an unavoidable regret at each round.  In the scenarios like the constructed hard instances of the proof, the identity of the rewarding action
encodes the user's private context. Consequently, DP limits how reliably the algorithm can select that rewarding action, leading to linear expected regret for fixed privacy parameters.   Moreover, the asymptotic competitive ratio in probability is at most $\delta$, which is typically chosen to be close to zero to guarantee effective privacy in real applications \cite{DP_FTRL_Federated_learning_xu2023federated,DP_online_learning_jain2012differentially,DP_MAB_cost_basu2019differential}. If it is required to preserve pure DP with $\delta=0$, no online allocation algorithm can achieve positive competitive ratio even asymptotically.  

\paragraph{Joint Differential Privacy}
In many practical scenarios, the standard $(\epsilon,\delta)$-DP is unnecessarily stringent because the action $x_t$ assigned to user $t$ is typically not directly observable by other users. For example, in battery-powered edge computing systems, a centralized server determines the resource allocation for each request without revealing these decisions to others. Each user can observe only the configuration of its own request and has no access to the allocations of other users.  It is also reasonable to assume that $x_t$ is observable only to the allocator and the user at round $t$ in many other examples (Section~\ref{sec:application}). Under this setting, it is not necessary to privatize the action output $x_t$ itself.

Nevertheless, privacy leakage can still arise through resource-based side channels that implicitly couple the actions of different users via shared resource constraints. Since users operate under a shared resource budget, the allocation decision for one user may reveal information of previous users through the remaining budget status or coordination variables. An adversary can send queries to the allocator acting as a normal user, thereby inferring sensitive information about others from its own observed allocations. 
To mitigate this risk, we adopt $(\epsilon,\delta)$-joint differential privacy (JDP, Definition~\ref{def:JDP}), which protects against privacy leakage induced by resource-constraint-based side channels. Compared to standard $(\epsilon,\delta)$-DP, $(\epsilon,\delta)$-JDP provides a meaningful relaxation by leveraging the fact that the action of a benign user $t$ is not observable by other users or attackers. This relaxed notion enables stronger reward guarantees while still preserving privacy against information leakage through shared resource constraints.
\begin{definition}[$(\epsilon, \delta)-$Joint DP (JDP)]\label{def:JDP}
For any $\epsilon\geq 0$, $\delta\in(0,1)$, an algorithm $\mathcal{A}: \mathcal{Y}^T\rightarrow \mathcal{X}^T$ is ($\epsilon, \delta$)-jointly differential private if for any two context sequences $\bm{y}, \bm{y}'\in \mathcal{Y}^T$ that differ in the input of a user $t$, and any subset $O\subset \mathcal{X}_{-t}$, $\Pr(\mathcal{A}(\bm{y})_{-t}\in O)\leq e^{\epsilon}\cdot \Pr(\mathcal{A}(\bm{y}')_{-t}\in O)+\delta$, where $\mathcal{X}_{-t}=\mathcal{X}_1\times \mathcal{X}_2\times \cdots\mathcal{X}_{t-1}\times \mathcal{X}_{t+1}\times\cdots  \mathcal{X}_{T}$ is the output space excluding the output space of user $t$ and $\mathcal{A}(\bm{y})_{-t}=\left[\mathcal{A}(\bm{y})_1, \mathcal{A}(\bm{y})_2, \cdots\mathcal{A}(\bm{y})_{t-1}, \mathcal{A}(\bm{y})_{t+1},\cdots,  \mathcal{A}(\bm{y})_{T}\right]$ is the algorithm output space excluding that of user $t$.
\end{definition}
An upper bound of the competitive ratio for online allocation with $(\epsilon, \delta)$-JDP is provided in the next Theorem. It demonstrates that the adversarial performance limits of online allocation degrades as the 
$(\epsilon,\delta)$-JDP requirement becomes more stringent (More discussions in Appendix~\ref{sec:challenges_JDP_LDP}).  
\begin{theorem}\label{thm:JDP_CR_upper_bound}
    For any $\epsilon>0, 0<\delta<1$, no $(\epsilon,\delta)-$JDP preserving online allocation algorithm $\mathcal{A}$ can achieve a competitive ratio $(\inf_{\bm{y}\in\mathcal{Y}^T} \frac{F_T^{\mathcal{A}}(\bm{y})}{\mathrm{OPT}(\bm{y})})$ higher than $1-\frac{1-\delta}{e^{\epsilon}+1}$. 
\end{theorem}

\paragraph{Local Differential Privacy} 
Both standard $(\epsilon,\delta)$-DP and $(\epsilon,\delta)$-JDP are designed for centralized settings, where a trusted allocator determines the resource allocation. However, in many scenarios (Section~\ref{sec:application}), allocation decisions are made locally at the user side. In this setting, each user does not share its raw allocation action with others, and thus there is no need to privatize the action itself. Nevertheless, users must share coordination signals (e.g., remaining budget states) to enforce globally shared resource constraints, which may still reveal sensitive information.

To protect privacy in this decentralized setting, we introduce local differential privacy (LDP) \cite{LDP_RL_garcelon2021local,LDP_bandits_zheng2020locally,LDP_DP_MDP_liao2023locally,LDP_robust_bandits_zhou2024locally} to regulate the sharing of coordination information. Without directly privatizing the action outputs, $(\epsilon,\delta)$-LDP provides a more relaxed notion than standard $(\epsilon,\delta)$-DP, allowing greater flexibility in achieving reward guarantees. Moreover, $(\epsilon,\delta)$-LDP implies $(\epsilon,\delta)$-JDP if the local mechanism $\mathcal{M}$ is equivalently operated by a centralized allocator, as formalized in Lemma~\ref{lma:LDP-JDP}.
\begin{definition}[$(\epsilon, \delta)-$Local DP ( LDP)]\label{def:LDP}
For any $\epsilon>0$, $\delta\in(0,1)$, a local privacy preserving mechanism $\mathcal{M}: \mathcal{Y}\rightarrow \mathcal{Z}$, where $\mathcal{Y}$ is the local context space and $\mathcal{Z}$ is the space of shared coordination information to other users, is ($\epsilon, \delta$)-locally differential private if for any adjacent contexts $y, y'\in \mathcal{Y}$, and any subset of $O\subset \mathcal{Z}$, $\Pr(\mathcal{M}(y)\in O)\leq e^{\epsilon}\cdot \Pr(\mathcal{M}(y')\in O)+\delta$.
\end{definition}

\begin{wrapfigure}[19]{R}{0.56\textwidth}
\vspace{-8mm}
\begin{minipage}{0.56\textwidth}
\begin{algorithm}[H]
\caption{\ouralg: Meta Algorithm for Privacy Preserving Online Allocation
}\label{alg:algorithm_meta}
	\begin{algorithmic}[1]
  \REQUIRE Initialize virtual budget  {$\tilde{B}_1=B_1=\rho T$}, dual variable $\mu_1$, and   step size $\eta>0$.
\FOR   {$t=1$ to $T$}
 \STATE 
 
 {\color{gray}\textit{Primal update (Line 3-4)}}
        \STATE Pre-select action:\\ $\hat{x}_t=\arg\max_{x\in\mathcal{X}_t} \left\{f_t(x)-\mu_t^\top b_t(x)\right\}$ 
      \STATE If $b_t(\hat{x}_{t})\leq \tilde{B}_t-B_{\mathrm{gap,t}}\mathbf{1}_m$, $x_t=\hat{x}_t$; Else, $x_t=0$

     {\color{gray}\textit{Private budget \& dual update (Line 5-6)}}
	\STATE Differential private remaining budget update:  \\
    $\tilde{B}_{t+1}=\mathcal{M}_B(\{b_\tau(x_{\tau})\}_{\tau\in[t]},\epsilon,\delta)$ (Joint DP)\\
or $\tilde{B}_{t+1}=\mathcal{M}_B(\tilde{B}_{t},b_t(x_t),\epsilon,\delta)$ (Local DP)
  \STATE Differential private dual update:\\
  $\mu_{t+1}=\mathcal{M}_D(\{b_\tau(x_{\tau})\}_{\tau\in[t]},\epsilon,\delta)$ (Joint DP)\\
  or   $\mu_{t+1}=\mathcal{M}_D(\mu_t, b_t(x_t),\epsilon,\delta)$ (Local DP)
  \ENDFOR
 \end{algorithmic}
\end{algorithm}
\end{minipage}
\end{wrapfigure}
Since $(\epsilon,\delta)$-LDP implies $(\epsilon,\delta)$-JDP when a 
locally privacy-preserving algorithm is equivalently implemented by a centralized 
allocator, as formalized in Lemma~\ref{lma:LDP-JDP}, online allocation 
under $(\epsilon,\delta)$-LDP does not admit a higher competitive ratio upper bound than $(\epsilon,\delta)$-JDP.

\section{Meta Algorithm}\label{sec:meta_alg}

\subsection{Meta Algorithm Design}
To preserve $(\epsilon,\delta)-$JDP or $(\epsilon,\delta)-$LDP, the meta algorithm in Algorithm~\ref{alg:algorithm_meta} performs an action selection (primal update) for user $t$ based on a clean context $y_t$, a privatized dual variable $\mu_t$, and a privatized remaining budget $\tilde{B}_t$.  
The action is pre-selected by balancing the immediate reward optimization with the budget saving for potentially higher future rewards with the privatized $\mu_t$ as the balancing weight, i.e.
\(   \hat{x}_t= \arg\max_{x\in\mathcal{X}_t} \left\{f_t(x)-\mu_t^\top b_t(x)\right\}.\)
Next in Line~4 of Algorithm~\ref{alg:algorithm_meta}, the allocator selects the action $x_t$ by projecting the pre-selected action $\hat{x}_t$ to ensure the budget constraint satisfaction. This projection must guarantee the budget consumption $b_t(x_t)$ at each round $t$ does not exceed the lower bound of the true remaining budget $B_t=B_1-\sum_{\tau=1}^{t-1}b_{\tau}(x_{\tau})$, which is estimated by subtracting the gap bound $B_{\mathrm{gap},t}$ between the privatized and clean budgets from the privatized remaining budget $\tilde{B}_t$.

To preserve $(\epsilon,\delta)-$JDP or $(\epsilon,\delta)-$LDP, Algorithm~\ref{alg:algorithm_meta} explicitly performs differential private mechanisms for the remaining budget and the Lagrangian dual as shown in Line 5-6 because the two variables are dependent on the context information of previous users.  
If we only require to preserve JDP, the mechanisms have access to all previous clean budget consumptions.  However, to preserve LDP, the remaining budget and the Lagrangian dual are updated locally and can only access the privatized variables from previous rounds.
Next, we present the conditions to guarantee the privacy notions, budget constraints, and allocation performance by \ouralg.

\subsection{Conditions for Privacy Preservation}
In online allocation, each action $x_t$ is a function of the clean context $y_t$ and the intermediate variables $\tilde{B}_t$ and $\mu_t$, so Billboard Lemma~\ref{lma:DP_JDP} directly implies the following theorem about the joint or local DP guarantees of Algorithm~\ref{alg:algorithm_meta} as indicated by the theorem below (proof in Appendix~\ref{sec:proof_meta_privacy}).
\begin{theorem}\label{thm:DP_meta}
If the compositions of the remaining budget $(\tilde{B}_1,\cdots, \tilde{B}_T)$ and the dual variables $(\mu_1,\cdots,\mu_T)$ satisfy $(\epsilon,\delta)-$DP,  Algorithm \ref{alg:algorithm_meta} satisfies $(\epsilon,\delta)-$JDP if it is implemented as a central privacy preserving algorithm, and $(\epsilon,\delta)-$LDP if it is implemented as a local privacy preserving algorithm. 
\end{theorem}
Theorem \ref{thm:DP_meta} shows that as long as the mechanisms $\mathcal{M}_B$ and $\mathcal{M}_D$ in Algorithm~\ref{alg:algorithm_meta} are $(\epsilon,\delta)-$differential private, then  Algorithm~\ref{alg:algorithm_meta} which maps user contexts to actions is $(\epsilon,\delta)-$ JDP or LDP.

\subsection{Conditions for Performance Guarantee}
Under the $(\epsilon,\delta)-$ JDP or LDP by Theorem~\ref{thm:DP_meta}, we derive the conditions to satisfy the budget constraints and guarantee asymptotic performance in both adversarial and i.i.d. stochastic settings. 

\begin{theorem}
\label{thm:reward_meta}
 The following statements hold for Algorithm~\ref{alg:algorithm_meta} with prescribed parameter $\zeta\in(0,1)$.   \\
(1) \textbf{Constraint feasibility.} Algorithm~\ref{alg:algorithm_meta} satisfies the constraint feasibility with high probability at least $1-\zeta$ as Definition~\ref{def:feasibility} if, for every fixed sequence $\bm y\in\mathcal Y^T$, it holds with probability at least $1-\zeta$ over the randomness of $\mathcal A$ that \(\forall t\in [T], \|\tilde B_t-B_t\|_\infty \leq B_{\mathrm{gap},t}\).\\
(2) \textbf{Performance guarantee.} Under the conditions in (1), Algorithm~\ref{alg:algorithm_meta} guarantees an asymptotic competitive
ratio of at least $1/\alpha$, $\alpha\geq 1$ in probability as Definition~\ref{def:cr} and a sublinear expected regret if  $B_{\mathrm{gap},t}\leq \Gamma(T,\zeta)$ and for every fixed sequence $\bm y\in\mathcal Y^T$,  with probability at least $1-\zeta$ over the randomness of $\mathcal A$, \( \forall \tau\in [T], \forall\mu\in\mathcal D, \sum_{t=1}^{\tau} \bigl(w_t(\mu_t)-w_t(\mu)\bigr) \leq \Gamma(T,\zeta)\), where $\Gamma(T,\zeta)$ is deterministic and the algorithm uses $\zeta=\zeta_T\in (0,1)$ satisfying \(\lim_{T\to\infty}\zeta_T=0, T\zeta_T=O(1), \lim_{T\to\infty}\frac{\Gamma(T,\zeta_T)}{T}=0\), and the dual region $\mathcal{D}\subset\mathcal{R}_+^m$ contains $\{\bm{0}, \frac{\bar{f}}{\alpha \rho_1}\mathbf{e}_1,\cdots, \frac{\bar{f}}{\alpha \rho_m}\mathbf{e}_m, \frac{\bar{f}}{\rho_1}\mathbf{e}_1,\cdots, \frac{\bar{f}}{\rho_m}\mathbf{e}_m\}$ where $\mathbf{e}_j\in \mathcal{R}^m$ has 1 at $j-$th entry and 0 at the others.
\end{theorem}
The proof is in Appendix~\ref{sec:proof_meta_reward}.
The asymptotic competitive ratio 
$\frac{1}{\alpha}$ in probability, achieved by Algorithm~\ref{alg:algorithm_meta} is optimal for online allocation with adversarial inputs, even in the absence of JDP or 
LDP requirements \cite{learning_auctions_balseiro2019learning}. Therefore, Theorem~\ref{thm:reward_meta} shows that, when its two 
conditions are satisfied, \ouralg asymptotically 
approaches the optimal performance in both adversarial and i.i.d. 
stochastic settings.
The first condition requires that the reserved budget cap $B_{\mathrm{gap},T}$ 
in Line~4 of Algorithm~\ref{alg:algorithm_meta} upper bounds the discrepancy between 
the privatized remaining budget and its clean counterpart, and increases in a sublinear way with $T$. This condition 
guarantees at each round the selected action satisfies the constraints with arbitrarily high probability. The second condition requires that the dual update in Line~6 
of Algorithm~\ref{alg:algorithm_meta} ensures convergence of the dual objective 
relative to any fixed dual variable in the set $\mathcal{D}$. The resulting  convergence rate directly 
determines the primal regret performance.

In practice, context distributions may vary over time, and contexts
may exhibit temporal dependence. The asymptotic competitive ratio
in probability of \ouralg established in Theorem~\ref{thm:reward_meta} continues to hold in this setting, since its failure-probability bound is uniform over all context sequences. However, sublinear expected regret is not guaranteed without additional assumptions, as discussed in Section~\ref{sec:non-stationary_discussions}.

By the conditions in Theorem~\ref{thm:DP_meta} and 
Theorem~\ref{thm:reward_meta}, the private online allocation can be converted
to a problem of designing differential private online learning algorithms with bounded convergence error. 
In this way, \ouralg provides a unifying interface for a wide range of existing and future privacy-preserving online learning designs to meet different privacy and performance requirements.

\section{Instantiated Algorithms and Performance Analysis}\label{sec:local_algrithm}
In this section, we instantiate \ouralg\ (Algorithm~\ref{alg:algorithm_meta}) to obtain \ouralgdmd, which preserves both JDP and LDP, and \ouralgftrl, which preserves only JDP, and provide performance analysis to characterize the privacy-performance trade-offs.

\subsection{ \ouralgdmd: Algorithm for JDP and LDP}\label{sec:online_allocation}\label{ppoa_dmd_algorithm}

\begin{wrapfigure}[12]{R}{0.56\textwidth}
\vspace{-3mm}
\begin{minipage}{0.56\textwidth}
\begin{algorithm}[H]
\caption{\ouralgdmd
}\label{alg:algorithm_local}
	\begin{algorithmic}[1]
 \REQUIRE Initialize virtual budget  {$\tilde{B}_1=B_1=\rho T$}, dual variable $\mu_1$, noise variance $\sigma^2$, step size $\eta>0$. 
\FOR   {$t=1$ to $T$}
      \STATE Select an action by Line 2-4 in Algorithm~\ref{alg:algorithm_meta}.
      \STATE \label{line:privatization} Privatize the budget consumption: \\$\tilde{b}_t(x_t)=b_t(x_{t})+n_t$, $n_t\sim \mathcal{N}(0,\sigma^2\mathbf{I})$.
	\STATE \label{line:budgetupdate} Update remaining budget $\tilde{B}_{t+1}=\tilde{B}_{t} -\tilde{b}_t(x_t)$.  
  \STATE Update the dual variable by private DMD in \eqref{eqn:DMD}. \label{line:dmd}
  \ENDFOR
 \end{algorithmic}
\end{algorithm}
\end{minipage}
\end{wrapfigure}
\ouralgdmd is presented in
Algorithm~\ref{alg:algorithm_local}, which can be implemented in either centralized or decentralized allocation systems. In a centralized system, each arriving user uploads its raw context to the centralized allocator, and the allocator selects an action for the user by Line 2 in Algorithm~\ref{alg:algorithm_local}. Next, to protect the user context information, the centralized allocator updates the remaining budget and the dual variable in a differential private way by Line 3-5 in Algorithm~\ref{alg:algorithm_local} and uses them to select action for the next user. 

In a decentralized system, each user at round $t$ makes the allocation decision locally by Line 2 in Algorithm~\ref{alg:algorithm_local} without uploading or sharing its raw context and selected action. Then, the user only disseminates the privatized budget consumption $\tilde{b}_t(x_t)$ in Line 3 of  Algorithm~\ref{alg:algorithm_local} to other users. The user at the next round $t+1$ updates the remaining budget and dual variable by Line 3-5 in Algorithm~\ref{alg:algorithm_local} based on the privatized consumption $\tilde{b}_t(x_t)$ and use them to select an action locally. 

Algorithm~\ref{alg:algorithm_local} needs a  dual learning procedure to update the dual variable in Line 5. Directly applying Dual Mirror Descent (DMD) in \cite{online_learning_orabona2019modern} updates the dual variable as \(\mu_{t+1}
=\arg\!\min_{\mu\geq 0}
(\rho-\tilde{b}_t(x_t))^\top \mu
+\frac{1}{\eta} V_h(\mu,\mu_t)\) where
\(
V_h(\mu,\mu_t)
= h(\mu)-h(\mu_t)-\nabla h(\mu_t)^{\top}(\mu-\mu_t)
\)
denotes the Bregman divergence associated with a $1$-strongly convex function
$h(\mu)$. However, noise in $\tilde{b}_t(x_t)$ accumulates in the DMD process, which induces a linear term in the convergence bound of the dual objective, violating Condition (2) in Theorem~\ref{thm:reward_meta}.  To this end, \ouralgdmd employs a \emph{projected} DMD update, i.e.
\begin{equation}\label{eqn:DMD}
\mu_{t+1}
=\arg\!\min_{\mu\geq 0,\ \|\mu\|_2\leq \mu_{\max}}
(\rho-\tilde{b}_t(x_t))^\top \mu
+\frac{1}{\eta} V_h(\mu,\mu_t),
\end{equation}
which constrains the dual variable norm by a finite
hyperparameter $\mu_{\max}>0$. This projection stabilizes the dual learning
process by avoiding unreasonably large dual variable due to unbounded noise injection, which is essential for preserving the worst-case performance. 

\paragraph{Analysis for \ouralgdmd}
Theorem \ref{thm:privacy_local} (proof in Appendix~\ref{sec:proof_JDP_local}) specifies the way to set the noise standard deviation $\sigma$ to preserve $(\epsilon,\delta)-$JDP and $(\epsilon,\delta)-$LDP. 
\begin{theorem}\label{thm:privacy_local}
Choosing $\sigma=\frac{\sqrt{2}\bar{b}(\sqrt{\ln(1/\delta)}+\sqrt{\ln(1/\delta)+\epsilon})}{\epsilon}$, \ouralgdmd (Algorithm \ref{alg:algorithm_local}) preserves $(\epsilon,\delta)-$LDP in decentralized settings and $(\epsilon,\delta)-$JDP in centralized settings. 
\end{theorem}

 Next, we prove that the privacy mechanism in \ouralgdmd satisfies the conditions in Theorem~\ref{thm:reward_meta} (proof in  Appendix~\ref{sec:proof_constraint_local}), thus asymptotically approaching the optimal solution for both i.i.d. stochastic and adversarial settings. 
\begin{theorem}\label{thm:reward_local}
 With $\sigma$ set in Theorem~\ref{thm:privacy_local}, the budget gap as $B_{\mathrm{gap},t}=\sigma \sqrt{2t\ln\frac{2 mT}{\zeta}}$, the step size as $\eta=\mathcal{O}(\frac{\epsilon}{\sqrt{T\ln(1/\delta)}})$, dual norm bound as $
\mu_{\max} \geq \max_j \frac{\bar{f}}{\rho_j}
$, and the failure probability as $\zeta=\zeta_T=\frac{1}{T^\vartheta+1}$ given a constant $\vartheta\geq 1$,
\ouralgdmd satisfies the conditions in Theorem~\ref{thm:reward_meta} with \(\Gamma(T,\zeta_T)=\widetilde{\mathcal{O}}\left(\frac{\mu_{\max}\sqrt{T\log(1/\delta)}}{\epsilon}\right)\) where $\widetilde{\mathcal{O}}$ absorbs the terms scaling with $\log(T)$, so it achieves an asymptotic competitive ratio in probability (Definition~\ref{def:cr}) of at least $1/\alpha$ with $\alpha=\sup_{j\in[m]}\frac{\bar{b}_j}{\rho_j}$ and a sublinear expected regret under i.i.d. arrivals in the order of \(\widetilde{\mathcal{O}}\left(\frac{\mu_{\max}\sqrt{T\log(1/\delta)}}{\epsilon}\right)\). 
\end{theorem}

Theorem~\ref{thm:reward_local} shows that \ouralgdmd with $(\epsilon,\delta)$-JDP/LDP has regret scaling with $\sqrt{\log(1/\delta)}/\epsilon$, illustrating a privacy-performance trade-off: higher DP levels (smaller $\epsilon$ or $\delta$) lead to larger regrets.  Furthermore, selecting a smaller $\mu_{\max}$ in projected DMD~\eqref{eqn:DMD} leads to smaller regret bounds since a smaller $\mu_{\max}$ reduces the impact of noise more effectively. However, $\mu_{\max}$ must not be overly small to ensure \ouralgdmd can search a proper dual variable. Theorem~\ref{thm:reward_meta} shows that $\mu_{\max}$ should be at least $\max_j \frac{\bar{f}}{\rho_j}$. This is because with less budget (smaller $\rho_j$) or larger reward bound $\bar{f}$, a proper dual variable must be larger to encourage resource saving or match the reward scaling.

\subsection{\ouralgftrl: Enhanced Algorithm for JDP}\label{sec:central_Algorithm}

In this section, we show that an enhanced instantiation of \ouralg can achieve a better privacy–performance trade-off when only JDP is required in a centralized setting. Intuitively, the advantage of the JDP-preserving centralized setting stems from the presence of a trusted allocator with access to users’ raw contexts. By \ouralg, the decision at each round does not depend on individual historical data, but rather on aggregated information from past users, such as the remaining budget and dual variables for resource coordination.
Unlike the LDP-preserving decentralized setting, where information aggregation is based on privatized local information, the centralized allocator can leverage raw data to construct more accurate aggregations. As a result, to achieve the same level of differential privacy, centralized mechanisms (e.g., \cite{DP_FTRL_kairouz2021practical}) require significantly less noise than LDP-based aggregation. Therefore, when only JDP is required, we aim to exploit centralized aggregation in \ouralg to enable substantially improved allocation performance.

Motivated by this goal,
we develop \ouralgftrl  (Algorithm~\ref{alg:algorithm_ftrl} in Appendix~\ref{sec:ppoa_ftrl_Algorithm}). 
The mirror descent update in \ouralgdmd depends on the dual update of the last round, so it cannot utilize the centralized information aggregation. Comparably, \ouralgftrl relies on Follow the Regularized Leader (FTRL) \cite{online_learning_orabona2019modern} to update the dual variable directly based on the aggregated budget consumption. It privatizes cumulative budget consumption
$\sum_{\tau=1}^{t} b_\tau(x_\tau)$ as $s_t$ by the tree aggregation strategy \cite{DP_FTRL_kairouz2021practical} which injects much less noise than directly perturbing the cumulative consumption to preserve the same level of $(\epsilon,\delta)$-JDP. Based on the privatized cumulative consumption $s_t$, we can replace Line 5-6 in Algorithm~\ref{alg:algorithm_meta} by the remaining budget update as
$\tilde{B}_{t+1}=B_1-s_t$ and the private dual update as
\(
\mu_{t+1}
=\arg\min_{\mu\geq 0}
(\rho t - s_t)^\top \mu
+\frac{1}{\eta} h(\mu),
\)
where $h(\mu)$ is a $1$-strongly convex regularization function. 
Further details about \ouralgftrl are provided in
Section~\ref{sec:ppoa_ftrl_Algorithm}.

\paragraph{Analysis for \ouralgftrl}
The $(\epsilon,\delta)-$JDP preservation with the selection of noise standard decision is justified below (proof in Appendix~\ref{sec:proof_JDP_central}). 
\begin{theorem}\label{thm:privacy_ftrl}
   \ouralgftrl (Algorithm \ref{alg:algorithm_ftrl}) preserves $(\epsilon,\delta)-$JDP if the standard deviation of the Gaussian noise added to each node of the tree is set as $\sigma=\frac{\sqrt{2}\bar{b}\sqrt{\lceil\log_2(T)\rceil}(\sqrt{\ln(1/\delta)}+\sqrt{\ln(1/\delta)+\epsilon})}{\epsilon}$.
\end{theorem}

Next, based on the conditions in Theorem~\ref{thm:reward_meta} and the convergence of FTRL with the tree aggregation mechanism, we can get the performance guarantee in the next theorem (proof in Appendix~\ref{sec:proof_central}). 
\begin{theorem}\label{thm:reward_central}
With $\sigma$ set in Theorem~\ref{thm:privacy_ftrl}, the budget gap as $B_{\mathrm{gap},t}=\sigma \sqrt{2\left\lceil\log_2(t)\right\rceil\ln\frac{2 mT}{\zeta}}$, the step size as $\eta=\mathcal{O}(\sqrt{\frac{\epsilon}{T\sqrt{\log_2(T)\ln(1/\delta)}}})$, and the failure probability as $\zeta=\zeta_T=\frac{1}{T^\vartheta+1}$ given a constant $\vartheta\geq 1$,
\ouralgftrl satisfies the conditions in Theorem~\ref{thm:reward_meta} with \(\Gamma(T,\zeta_T)=\widetilde{\mathcal{O}}\left(\frac{\sqrt{T\sqrt{\log(1/\delta)}}}{\sqrt{\epsilon}}\right)\) where $\widetilde{\mathcal{O}}$ absorbs the terms scaling with $\log(T)$, so it achieves an asymptotic competitive ratio in probability (Definition~\ref{def:cr}) of at least $1/\alpha$ with $\alpha=\sup_{j\in[m]}\frac{\bar{b}_j}{\rho_j}$ and a sublinear expected regret under i.i.d. arrivals in the order of \(\widetilde{\mathcal{O}}\left(\frac{\sqrt{T\sqrt{\log(1/\delta)}}}{\sqrt{\epsilon}}\right)\).  
\end{theorem}
We can find that the regret bounds of
\ouralgftrl scale with $\widetilde{\mathcal{O}}(\sqrt{\sqrt{\log(1/\delta)}/\epsilon})$ which is the squared root of the privacy-related scaling  (Theorem~\ref{thm:reward_local}) achieved by \ouralgdmd. This indicates that within the regime of stringent privacy requirements (e.g. $\epsilon<1$ and $\delta$ is close to zero), \ouralgftrl provides significantly better allocation performance than \ouralgdmd. This benefit stems from the FTRL-based aggregation,
which introduces much less Gaussian noise to the aggregated information.

\section{Numerical Results}\label{sec:experiments}
We consider a request routing problem for a battery-powered edge AI system, which enables time-sensitive edge AI services \cite{Multi-agent-router_yue2025masrouter,edge_AI_service_placement_lin2021optimizing,edge_survey_luo2021resource,AI_edge_computing_tuli2023ai}.  Beyond this, the numerical evaluation on the application of online biding in advertising \cite{liao2014ipinyou,bidding_auction_online_advertising_aggarwal2024auto} is provided in Appendix~\ref{sec:more_application}.
The edge AI service can deploy multiple models on an edge data center with finite energy budget in its battery. As user requests arrive sequentially, the system must decide in real time which model should be selected to serve the request. The experiments are based on the RouterBench dataset \cite{Routerbench_hu2024routerbench} and setup details are given in the Appendix~\ref{sec:experimentsetups}. 

\textbf{Baselines}. 
We compare \ouralgdmd and \ouralgftrl against the following baselines. \textbf{Non-private baselines}: \textbullet\; \textsf{\greedy}: greedy selects
the constraint-feasible action that directly maximizes the reward $f_t(x)$. \textbullet\; \textsf{\dmd}: the dual mirror descent algorithm \cite{dual_mirror_descent_online_allocation_balseiro2020dual} with $h(\mu)$ being the squared norm function.
\textbullet\;\textsf{\ftrl}: update the dual variable with follow the regularized leader \cite{online_learning_orabona2019modern} with $h(\mu)$ being the squared norm function.
\textbf{Private baselines}: \textbullet\;\textsf{\oadp}: a centralized online allocation algorithm preserving standard $(\epsilon,\delta)-$DP in Definition~\ref{def:DP} by additionally applying Gaussian mechanism to reward and consumption functions of \ouralgftrl. \textbullet\; \textsf{\ndmd}: Noisy DMD \cite{noisy_DMD_chen2024noisy} adapted to satisfy the constraint feasibility in the same way as \ouralg without applying the projection in \ref{eqn:DMD}. \textbullet\;\textsf{\nftrl}: Directly applying noisy FTRL without applying tree aggregation (TA) \cite{tree_aggredation_chan2011private} for consumption aggregation. \textbullet\;\textsf{\pridown}: Algorithm 2 in \cite{JDP_packing_huang2018near} adapted to satisfy the constraint feasibility in the same way as \ouralg.

\begin{table*}[!t]
\scriptsize
\vspace{-0.2cm}
\centering
\resizebox{\textwidth}{!}{
\begin{tabular}{c|c c c| c c c c c c } 
\toprule
\multirow{2}{*}{\textbf{Algorithms}} & \multicolumn{3}{c|}{Non-private algorithms} & \multicolumn{5}{c}{Private algorithms}\\ 
    & \textsf{\greedy} & \textsf{\dmd} & \textsf{\ftrl} & \textsf{\oadp} & \textsf{\ndmd} & \textsf{\nftrl} & \pridown & \textbf{\ouralgdmd} & \textbf{\ouralgftrl}   \\ 
\midrule
AVG  & 0.710&\textbf{0.919} & 0.914 & 0.591$\pm$0.0001& 0.778$\pm$0.001  &0.726$\pm$0.001 &0.775$\pm$ 0.0006 & 0.801$\pm$0.001 & \textbf{0.879}$\pm$0.0002 \\
CR   & 0.632 & \textbf{0.876} & 0.867  & 0.514$\pm$0.003 &  0.635$\pm$0.015 & 0.597$\pm$0.019 & 0.652$\pm$0.022 &0.665$\pm$0.023 &\textbf{0.834}$\pm$0.005  \\
\bottomrule
\end{tabular}
}
\caption{\small{The expected (AVG) and adversarial (CR) performance of algorithms. Among the private algorithms, \oadp admits $(\epsilon,\delta)-$DP, \ndmd and \ouralgdmd admit $(\epsilon,\delta)-$JDP or $(\epsilon,\delta)-$LDP, \nftrl and \ouralgftrl admit $(\epsilon,\delta)-$JDP, all with $\epsilon=10$ and $\delta=0.01$.}}
\label{table:comparison_all_results}
\end{table*}

\paragraph{Baseline Comparison}
We evaluate both the expected and adversarial performance of the algorithms in Table~\ref{table:comparison_all_results}. All methods exhibit zero budget violation on the dataset, ensuring a fair comparison. As expected, \dmd\ and \ftrl\ achieve the best performance without privacy constraints. More importantly, when requiring both JDP and LDP, \ouralgdmd consistently outperforms the direct method \ndmd, demonstrating the effectiveness of the dual projection in~\eqref{eqn:DMD} for mitigating the impact of noise in the dual learning process. When only JDP is required, \ouralgftrl achieves significantly better performance than \nftrl, \ndmd, \pridown, and \ouralgdmd, as its tree-aggregation mechanism more effectively leverages raw historical contexts. Finally, \oadp enforces the strongest DP notion, so despite operating in a centralized way, it exhibits the worst performance among the private baselines, which empirically supports the impossibility result in Theorem~\ref{thm:DP_lower_bound}.

\begin{wrapfigure}[9]{r}{0.32\textwidth}
\vspace{-8mm}
    \centering
    \includegraphics[width=0.32\textwidth]{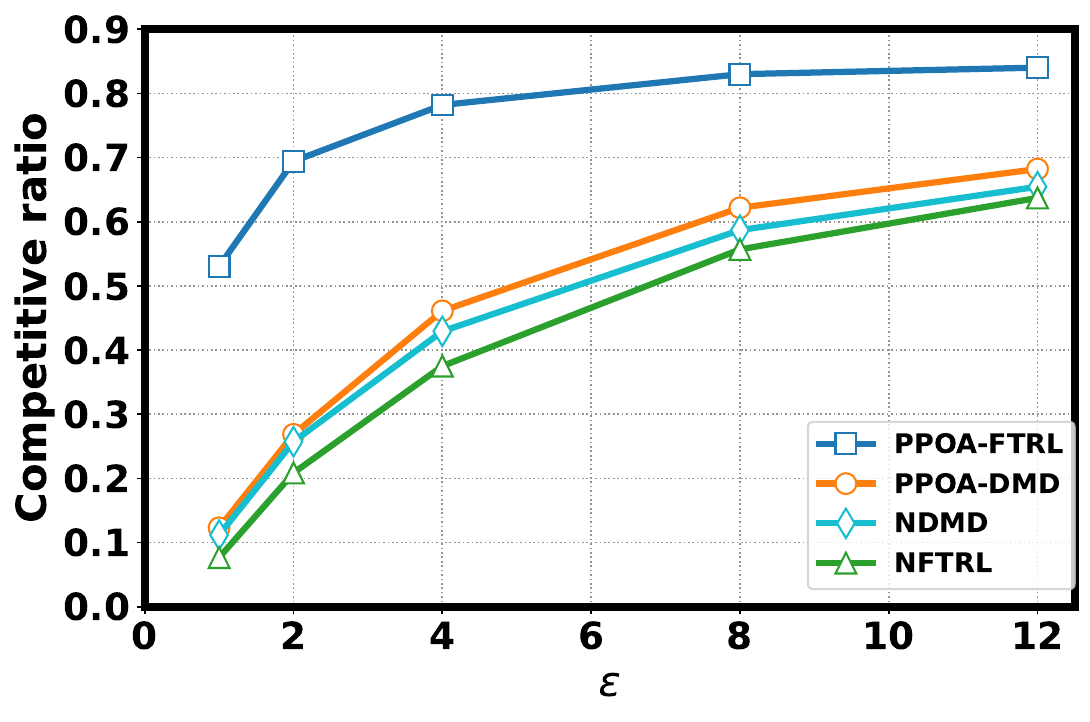}
    \vspace{-7mm}
    \caption{Competitive ratio with different $\epsilon$ under $\delta=0.01$.}
    \label{fig:trade-off_epsilon}
\end{wrapfigure}
\paragraph{Privacy-Performance Trade-off}
Figure~\ref{fig:trade-off_epsilon} reports the competitive ratio as a function of the privacy level~$\epsilon$, illustrating the privacy–performance trade-off. As expected, looser privacy requirements (larger~$\epsilon$) lead to improved reward performance across all methods. More importantly, among algorithms that satisfy both JDP and LDP, \ouralgdmd achieves a consistently better trade-off curve than \ndmd, owing to the dual projection in~\eqref{eqn:DMD}, which stabilizes the dual updates under noise injection. Furthermore, when only JDP is required, \ouralgftrl attains a significantly better trade-off than the other algorithms, highlighting the effectiveness of the tree-aggregation strategy for online allocation.

Results of ablation studies on privacy levels, budget reservation setups in \ouralg, dual norm cap in \ouralgdmd, and budget $B$ are available in Appendix~\ref{sec:additionalresults}.

\section{Conclusion}\label{sec:conclusion}
In this paper, we study the problem of online allocation with DP.
We first formally analyze the performance limits for different DP definitions.
We then propose a meta-algorithm~\ouralg  and establish sufficient conditions under which \ouralg preserves JDP and LDP, satisfies the budget constraint with any prescribed high probability, and achieves asymptotically optimal performance.
We instantiate \ouralg to JDP/LDP-preserving \ouralgdmd and an enhanced JDP-preserving \ouralgftrl with concrete performance analysis. 
Finally, we validate our theoretical findings through numerical studies. \\
\textbf{Limitations and future work.} While the work establishes the first performance bounds for online allocation under DP definitions, they are not proved to be the best bounds in terms of the scaling w.r.t. privacy levels. There is still room to improve the bounds especially in an almost-sure asymptotic or a non-asymptotic way. Also, it is interesting to explore meaningful performance guarantee under the most strict DP notion. \\
\textbf{Broader impacts.} The results in this work provide a algorithmic foundation to improve the privacy protection in wide online resource allocation systems. 

\section*{Acknowledgement}
We thank the NeurIPS reviewers for their insightful comments during the rebuttal process, which have helped us improve our work. Jianyi Yang was supported in part by the University of Houston Energy Transition Institute Seed Grant Program. Shaolei Ren was supported in part by the NSF under grants NSF-2449990 and NSF-2326598. Xingyu Zhou is supproted in part by the NSF under grants NSF CAREER-2441519 and NSF-2312835.

\newpage
\appendix
\section{Proofs and Discussions on Performance Limits of DP Definitions}\label{sec:lower_bounds}\label{sec:proofs_lower_bound}
\subsection{Proof of Theorem \ref{thm:DP_lower_bound}}

\paragraph{Lower bound of expected regret}
We prove the theorem by establishing an example below. 
Select a constant $K\geq 2$.  At every round $t$, the action set is $\mathcal X=\{0,1,\ldots,K\}$.
Let $S_t$ be a random variable uniformly distributed over $\{1,\ldots,K\}$ and $S_1,\cdots,S_T$ are independent and identically distributed. 
The reward and cost functions are defined as 
\begin{equation}\label{eq:hard-instance}
f_t(x)=\mathbf{1}\{x=S_t\},\qquad
b_t(x)=b\cdot\mathbf{1}\{x=S_t\},
\end{equation}
where $b\geq \rho$ is a constant. 
The total budget is $B=\rho \cdot T$ with $\rho\in (0,1)$. In the constructed example, the null action $0\in\mathcal X$ has zero reward and
cost, i.e. $f_t(0)=0,b_t(0)=0$. The rewards and costs are non-negative and upper bounded by one. Thus, the constructed example satisfies the considered problem setting in Section~\ref{sec:formulation}.

Let
\(
n_T=\left\lfloor\frac{\rho T}{b}\right\rfloor.
\)
A feasible policy chooses the rewarding action $x_t=S_t$ for the first $n_T$ rounds and the null action $x_t=0$ thereafter, obtaining total reward $n_T$. Since every unit of reward costs exactly $b$, any solution to the problem earns at most $n_T$ units of reward. Consequently, for any context sequence $\bm{y}$ in the construction, the optimal reward is
\begin{equation}\label{eqn:optimal_reward_example}
    OPT(\bm y)=n_T=\left\lfloor\frac{\rho T}{b}\right\rfloor.
\end{equation}

Next, we bound the expected reward given an algorithm $\mathcal{A}$ with DP guarantee.  Denote the probability that the algorithm $\mathcal{A}$ selects the $i-$th action ($i\in\{0,1,\ldots,K\}$) given $S_t=j$ ($j\in\{1,\ldots,K\}$) at round $t$ as
\[
p_{t,i\mid j}=\Pr(x_t=i\mid S_t=j) = \mathbb{E}_{\Pr(S_{-t})}\left[\Pr(x_t=i\mid S_t=j, S_{-t}) \right],
\]
where these probabilities average over the algorithm's internal randomness
and the contexts at all other rounds $S_{-t}=\{S_1,\cdots, S_{t-1}, S_{t+1}, S_T\}$. Due to the independence of the contexts across time, $\forall j\in\{1,\ldots,K\}, \Pr(S_{-t}\mid S_t=j)=\Pr(S_{-t})$.  Given two sequences that differ only in user $t$'s context $S_t=i$ and $S_t=j$, applying DP (Definition~\ref{def:DP}) to the observation of $x_t$ and averaging over the context sequence $S_{-t}$, it holds that
\begin{equation}\label{eq:pairwise-dp}
p_{t,i\mid i}\leq e^\epsilon p_{t,i\mid j}+\delta,
\qquad \forall i\neq j.
\end{equation}
Summing \eqref{eq:pairwise-dp} over all pairs $(i,j)$ with $i,j\in\{1,\ldots,K\}$ and $i\neq j$, we have
\begin{equation}\label{eqn:sum_rewarding_probability}
\begin{split}
(K-1)\sum_{i=1}^K p_{t,i\mid i}
&\leq e^\epsilon
 \sum_{j=1}^K\sum_{i=1,i\neq j}^K
p_{t,i\mid j}+K(K-1)\delta\\
&\leq e^\epsilon\sum_{j=1}^K(1-p_{t,j\mid j})+K(K-1)\delta\\
&=e^\epsilon\left(K-\sum_{i=1}^Kp_{t,i\mid i}\right)
 +K(K-1)\delta,
 \end{split}
\end{equation}
where the second inequality holds because $\sum_{i=1,i\neq j}^K p_{t,i\mid j}=1-p_{t,j\mid j}-p_{t,0\mid j}\leq 1-p_{t,j\mid j}$.
Since $S_t$ is uniformly drawn from $\{1,\ldots,K\}$, following \eqref{eqn:sum_rewarding_probability}, we have 
\begin{equation}\label{eqn:reward_upper_bound}
\mathbb E_{\mathcal A,\bm y}[f_t(x_t)] =\frac{1}{K}\sum_{i=1}^Kp_{t,i\mid i}
\leq \frac{e^{\epsilon}+(K-1)\delta}{e^{\epsilon}+K-1}\coloneqq q_K.
\end{equation}
Thus, the total expected reward is bounded as
\begin{equation}\label{eq:total-reward-upper}
\mathbb E_{\mathcal A,\bm y}[F_T^{\mathcal A}(\bm y)]\leq q_KT.
\end{equation}

Thus, the expected regret is lower bounded as \[
\mathbb E_{\mathcal A,\bm y}
\left[OPT(\bm y)-F_T^{\mathcal A}(\bm y)\right]\geq \left(\frac{\rho}{b}-q_K\right)T-1
\]
Since $q_K=\frac{e^\epsilon+(K-1)\delta}{e^\epsilon+K-1}
=\delta+\frac{e^\epsilon(1-\delta)}{e^\epsilon+K-1}$ and $K\geq 2$, it holds that $\delta \leq q_K< 1$, and $q_K\rightarrow\delta$ when the action set size approaches infinite ($K\rightarrow \infty$). Thus, there exist instances with $\frac{\rho}{b}>q_K$ such that the expected regret increases at least linearly, i.e. $\mathbb E_{\mathcal A,\bm y}
\left[OPT(\bm y)-F_T^{\mathcal A}(\bm y)\right]\geq \kappa T$ given a fixed $\kappa>0$. Thus, no algorithm can simultaneously guarantee DP and sublinear expected regret. 

\paragraph{Upper bound of asymptomatic competitive ratio}
Next, we prove the upper bound of asymptomatic competitive ratio in probability by contradiction using the hard example in \eqref{eq:hard-instance}.  If the there exists an algorithm with DP guarantee and a asymptomatic competitive ratio in probability $\frac{1}{\alpha}>\delta$, then by Definition~\ref{def:cr}, it holds for every $\kappa>0$ that,
\begin{equation}\label{eqn:cr_inprobability_copy}
\lim_{T\to\infty}
\sup_{\bm y\in\mathcal Y^T}
\Pr_{\mathcal A}
\left[ \frac{\mathrm{OPT}(\bm y) -\alpha F_T^{\mathcal A}(\bm y)}{T}>\kappa\right]=0.
\end{equation}

Since $\lim_{K\rightarrow \infty } q_K\to\delta<\frac{1}{\alpha}$ by the definition of $q_K$ in \eqref{eqn:reward_upper_bound}, we can find instances with finite $K$ such that $\alpha q_K<1$.
Choose the rewarding cost in the constructed instance as
\( b = \min\left\{ 1,\frac{2\rho}{1+\alpha q_K} \right\}.
\)
This construction ensures that
\(
\rho<b\leq1,  \frac{\rho}{ b} \geq\frac{1+\alpha q_K}{2}.
\)
Denote $\kappa'=\frac{1-\alpha q_K}{2}>0$. Following \eqref{eqn:optimal_reward_example} and \eqref{eq:total-reward-upper}, it holds for any $\alpha>0$ that 
\begin{equation}
\begin{split}
\mathbb E_{\mathcal A,\bm y}
\left[OPT(\bm y)-\alpha F_T^{\mathcal A}(\bm y)\right]
&\geq\left\lfloor\frac{\rho T}{b}\right\rfloor
 -\alpha q_K T \\
 &\geq\left(\frac{\rho}{b}-\alpha q_K\right)T-1\\
 &\geq \frac{1-\alpha q_K}{2}T-1= \kappa'T-1.
\end{split}
\end{equation}

Since rewards are nonnegative, we have
\(
\mathrm{OPT}(\bm y)-\alpha F_T^{\mathcal A}(\bm y)\leq \mathrm{OPT}(\bm y)\leq T
\)
for every sequence $\bm y$. Therefore, for every $\kappa>0$, it holds that
\[
\mathrm{OPT}(\bm y)
-\alpha F_T^{\mathcal A}(\bm y)\leq
\kappa T+ T\cdot \mathbf{1}\left\{ \mathrm{OPT}(\bm y) -\alpha F_T^{\mathcal A}(\bm y)>\kappa T\right\}.
\]
Taking expectations over the algorithm $\mathcal A$ and the sequence $\bm y$, we have
\[
\kappa' T-1\leq \mathbb E_{\mathcal A,\bm y}
\left[OPT(\bm y)-\alpha F_T^{\mathcal A}(\bm y)\right]
\leq
\kappa T+
T\Pr_{\mathcal A,\bm y}
\left[
\mathrm{OPT}(\bm y)
-\alpha F_T^{\mathcal A}(\bm y)>\kappa T
\right].
\]
Choosing the tolerance in \eqref{eqn:cr_inprobability_copy} as  $\kappa=\kappa'/2$, we have
\begin{equation}\label{eqn:contradiction_inequality}
\begin{split}
\lim_{T\to\infty}\sup_{\bm y\in\mathcal Y^T}
\Pr_{\mathcal A}
&\left[
\frac{\mathrm{OPT}(\bm y)
-\alpha F_T^{\mathcal A}(\bm y)}{T}
>\frac{\kappa'}{2}
\right]\geq \lim_{T\to\infty}\Pr_{\mathcal A,\bm y}
\left[
\frac{\mathrm{OPT}(\bm y)
-\alpha F_T^{\mathcal A}(\bm y)}{T}
>\frac{\kappa'}{2}
\right]\\
&=\lim_{T\to\infty}\Pr_{\mathcal A,\bm y}
\left[
\mathrm{OPT}(\bm y)
-\alpha F_T^{\mathcal A}(\bm y)>\kappa T
\right]\geq \lim_{T\to\infty}(\frac{\kappa'}{2}-\frac{1}{T})=\frac{\kappa'}{2}>0,
\end{split}
\end{equation}
where the first inequality holds since the supremum over inputs is at least the average over their distribution. 

Therefore, \eqref{eqn:contradiction_inequality} contradicts with the definition of asymptomatic competitive ratio in probability of at least $1/\alpha$ in Definition~\ref{def:cr}, and the asymptomatic competitive ratio in probability of any algorithm is at most $\delta$.

\subsection{Proof of Theorem \ref{thm:JDP_CR_upper_bound}}\label{sec:proof_JDP_CR_bound}
\begin{proof}
    We create a scenario to establish the upper bound of the competitive ratio as below. The considered scenario only has two rounds, and the action set for each round is $\mathcal{X}=\{0,1\}$. We assume the budget consumption function is $b_t(x)=x$ and the total budget is $B=1$. The scenario has two possible reward sequences which differ on the reward of the first round. The first sequence $\mathcal{S}_1$ is $f_1(x)=(1+k)\Delta\mathbb{I}(x), f_2(x)=\Delta\mathbb{I}(x)$ and occurs with a probability of $\rho$. The second sequence $\mathcal{S}_2$ is $f_1=0, f_2(x)=\Delta\mathbb{I}(x)$ and occurs with a probability of $1-\rho$. 

    Due to the budget constraint in the constructed scenario, the  feasible action sequences includes $\bm{x}^{(1)}=\{x_1=1,x_2=0\}$, $\bm{x}^{(2)}=\{x_1=0,x_2=0\}$ and $\bm{x}^{(3)}=\{x_1=0,x_2=1\}$. We can calculate the competitive ratios of all three feasible action sequences. Let the probability of $\mathcal{S}_1$ be $\Pr(\mathcal{S}_1)=p$ and the probability of $\mathcal{S}_2$ be $\Pr(\mathcal{S}_2)=1-p$,  the expected competitive ratio can be expressed as
    \begin{equation}
    \begin{split}
        \mathbb{E}_{\bm{y}\sim\Pr_{1:T}}&\bigg[\frac{F_T^{\mathcal{A}}(\bm{y})}{\mathrm{OPT}(\bm{y})}\bigg]=p \left( \Pr(\bm{x}^{(1)}\mid \mathcal{S}_1)+\frac{1}{1+k}\Pr(\bm{x}^{(3)}\mid \mathcal{S}_1)\right)+(1-p)\Pr(\bm{x}^{(3)}\mid \mathcal{S}_2)\\
        &= p \Pr(\bm{x}^{(1)}\mid \mathcal{S}_1)+\frac{p}{1+k}\bigg(1-\Pr(x_2=0\mid \mathcal{S}_1)\bigg)+(1-p)\bigg(1-\Pr(x_2=0\mid \mathcal{S}_2)\bigg)\\
        &\leq 1-\frac{kp}{1+k}+\frac{kp}{1+k}\Pr(x_2=0\mid \mathcal{S}_1)-(1-p) \Pr(x_2=0\mid \mathcal{S}_2)\\
        &\leq 1-\frac{kp}{1+k}+\frac{kp}{1+k}\Pr(x_2=0\mid \mathcal{S}_1)-(1-p) \left(e^{-\epsilon}\Pr(x_2=0\mid \mathcal{S}_1)-e^{-\epsilon}\delta\right)\\
        &\leq 1-\frac{kp}{1+k}+(1-p)e^{-\epsilon}\delta,
        \end{split}
    \end{equation}
    where the first inequality holds because $\Pr(\bm{x}^{(1)}\mid \mathcal{S}_1)\leq \Pr(x_2=0\mid \mathcal{S}_1)$, the second inequality holds by the definition of $(\epsilon,\delta)-$JDP, and last inequality holds by choosing $p\leq \frac{e^{-\epsilon}}{k/(1+k)+e^{-\epsilon}}$ and the fact $\Pr(x_2=0\mid \mathcal{S}_1)\in [0,1]$. 
    
 Choosing $p = \frac{e^{-\epsilon}}{k/(1+k)+e^{-\epsilon}}$ which minimizes the competitive ratio, the upper bound of the expected competitive ratio is $1-\frac{1-\delta}{e^{\epsilon}+1+\frac{1}{k}}$, which approaches $1-\frac{1-\delta}{e^{\epsilon}+1}$ when $k\rightarrow \infty$. Thus, $1-\frac{1-\delta}{e^{\epsilon}+1}$ is the upper bound of competitive ratio in the worst case. 
\end{proof}

\subsection{Discussions on the Performance Limits of JDP or LDP}\label{sec:challenges_JDP_LDP}
Theorem~\ref{thm:JDP_CR_upper_bound} provides a competitive ratio upper bound for any online allocation algorithm that satisfies $(\epsilon,\delta)-$JDP. The same upper bound also applies to any online allocation algorithm that satisfies $(\epsilon,\delta)-$LDP. The reason is that $(\epsilon,\delta)$-LDP implies $(\epsilon,\delta)$-JDP when a 
locally privacy-preserving algorithm is equivalently implemented by a centralized 
allocator, as formalized in Lemma~\ref{lma:LDP-JDP}. Consequently, online allocation 
under $(\epsilon,\delta)$-LDP cannot admits a better competitive ratio upper bound than $(\epsilon,\delta)$-JDP. 

In fact,
$(\epsilon,\delta)$-JDP (Definition~\ref{def:JDP}) 
and $(\epsilon,\delta)$-LDP (Definition~\ref{def:LDP}) 
provide privacy guarantees for central and local privacy-preserving models,  respectively, while being more relaxed than standard $(\epsilon,\delta)$-DP. 
JDP and LDP allow online algorithms greater flexibility in selecting actions in response to users’ contexts. Thus, while standard $(\epsilon,\delta)$-DP cannot guarantee any competitive ratio larger than $\delta$ which is close to 0, it is possible to guarantee meaningful performance bounds for online allocation with JDP or LDP. Interestingly, even as
$\epsilon$ and $\delta$ approach zero, the competitive ratio upper bound 
remains $\frac{1}{2}$, which matches the competitive ratio upper bound of a greedy algorithm that selects action at each round by maximizing the reward \cite{online_matching_mehta2013online}. 

Nevertheless, a fundamental tension between privacy preservation 
and allocation performance persists due to resource-constraint-based side channels: higher privacy levels can lead to lower competitive ratio upper bound (Theorem~\ref{thm:JDP_CR_upper_bound}).
This fundamental tension is illustrated by the JDP-preserving online allocation example in the proof of Theorem~\ref{thm:JDP_CR_upper_bound} in Section~\ref{sec:proof_JDP_CR_bound}.
The example is understood as below. The optimal strategy under the first situation $\mathcal{S}_1$ is to select $x_1 = 1$. In contrast, under the second situation $\mathcal{S}_2$, 
the optimal strategy is to select $x_1 = 0$ and reserve the budget to select 
$x_2 = 1$. JDP requires protecting the reward information of $f_1$ even when the second action $x_2$ is observable. However,  if an attacker observes $x_2 = 1$, they can infer that $x_1 = 0$ due to the budget constraint, which implies 
a high probability that the underlying situation is $\mathcal{S}_2$ 
(i.e., $f_1(x) = 0$). This privacy leakage arises from a resource-constraint-based side channel. To mitigate such leakage, a 
JDP-preserving algorithm is forced to adopt suboptimal strategies, thereby sacrificing allocation performance. 
This example establishes a competitive ratio upper bound for $(\epsilon, \delta)$-JDP preserving online allocation in Theorem~\ref{thm:JDP_CR_upper_bound}.

\section{More Details about \ouralgftrl }\label{sec:ppoa_ftrl_Algorithm}

\begin{wrapfigure}[16]{R}{0.56\textwidth}
\vspace{-5mm}
\begin{minipage}{0.56\textwidth}
\begin{algorithm}[H]
\caption{\ouralgftrl
}\label{alg:algorithm_ftrl}
	\begin{algorithmic}[1]
  \REQUIRE Initialize virtual budget  {$\tilde{B}_1=B_1=\rho T$}, dual variable $\mu_1$, noise variance $\sigma^2$, step size $\eta>0$. 
  \STATE Initialize a binary tree $\mathcal{T}\leftarrow \mathsf{InitializeTree}(T,\sigma^2)$. 
\FOR   {$t=1$ to $T$}
  \STATE Select an action by Line 2-4 in Algorithm~\ref{alg:algorithm_meta}. 
      \STATE Add new budget consumption to the tree: $\mathcal{T}\leftarrow \mathsf{AddToTree}(\mathcal{T},t, b_t(x_t))$.
      \STATE Get privatized sum of the consumption via Tree Aggregation: $s_t=\mathsf{GetSum}(\mathcal{T},t)$.
	\STATE Update remaining budget $\tilde{B}_{t+1}=B_1 -s_t$.  
  \STATE Update the dual variable by private FTRL in Eqn.~\eqref{eqn:DP-FTRL}.
  \ENDFOR
 \end{algorithmic}
\end{algorithm}
\end{minipage}
\end{wrapfigure}

\ouralgftrl is provided in Algorithm~\ref{alg:algorithm_ftrl}. \ouralgftrl makes primal decisions in the same
manner as Lines~2--4 of Algorithm~\ref{alg:algorithm_meta}. It then performs
privacy-preserving updates of the remaining budget and the dual variables.
In the non-private setting, a centralized online learning method based on
\emph{Follow the Regularized Leader} (FTRL) \cite{online_learning_orabona2019modern} can be applied to update the dual
variable by minimizing the regularized cumulative dual objective:
\(
\mu_{t+1}
=\arg\min_{\mu>0}
\left( \sum_{\tau=1}^{t} w_\tau(\mu) \right)
+\frac{1}{\eta} h(\mu),
\)
where $w_\tau(\mu)=(\rho-b_\tau(x_\tau))^\top \mu$ denotes the dual objective at
round $\tau$, and $h(\mu)$ is a $1$-strongly convex regularization function.
We find that both the FTRL dual update and the remaining budget update
$B_{t+1}=B_1-\sum_{\tau=1}^{t} b_\tau(x_\tau)$ depend on previous users' information through the cumulative budget consumption
$\sum_{\tau=1}^{t} b_\tau(x_\tau)$.
Thus, we privatize this consumption sum as $s_t$ in Line~5 of Algorithm~\ref{alg:algorithm_meta} using the tree-aggregation
mechanism \cite{DP_FTRL_kairouz2021practical}.
Substituting $s_t$ for the clean cumulative consumption yields the privatized dual update
\begin{equation}\label{eqn:DP-FTRL}
\mu_{t+1}
=\arg\min_{\mu\geq 0}
(\rho t - s_t)^\top \mu
+\frac{1}{\eta} h(\mu),
\end{equation}
and the privatized remaining budget is
$\tilde{B}_{t+1}=B_1-s_t$.

The tree-aggregation mechanism efficiently leverages clean historical information
to compute a privatized cumulative budget consumption while avoiding redundant
noise injection. In Line~1 of Algorithm~\ref{alg:algorithm_ftrl}, the function
$\mathsf{InitializeTree}(T,\sigma^2)$ constructs a binary tree
$\mathcal{T}$ with $2^{\lceil \log T \rceil}$ leaf nodes. Each node in the tree
is independently initialized by sampling from a Gaussian distribution
$\mathcal{N}(0,\sigma^2 \bm{I})$, which ensures that the value stored at every node
of $\mathcal{T}$ is differentially private.
After selecting the action $x_t$ at round $t$, the function
$\mathsf{AddToTree}(\mathcal{T}, t, b_t(x_t))$ adds the resource consumption
$b_t(x_t)$ to all nodes along the path from the $t$-th leaf node to the root of $\mathcal{T}$. The privatized cumulative budget consumption up to round $t$ is then obtained via the function $\mathsf{GetSum}(\mathcal{T}, t)$, which aggregates
the minimum number of tree nodes required to represent the prefix sum. This design
ensures that each update contributes to at most $\mathcal{O}(\lceil\log_2 t \rceil)$ noisy nodes.

The details of Tree-based Aggregation are provided below.
The key functions of the tree-based aggregation scheme \cite{DP_FTRL_kairouz2021practical} are detailed below.

1. InitializeTree($T,\sigma^2$): Initialize a binary tree $\mathcal{T}$ with $2^{\left \lceil \log T \right \rceil }$ leaf nodes.  Each node in the tree is sampled i.i.d. from $\mathcal{N}(0, \sigma^2\cdot \bm{I}_{m\times m})$.

2. AddToTree($\mathcal{T},t,b_t$): Add the resource consumption $b_t(x_t)$ to the values of all the nodes on the path starting from the $t-$th leaf node to the root of $\mathcal{T}$. 

3. GetSum($\mathcal{T},t$):  
Let $\{u_1,u_2,\cdots, u_h\}$ be the nodes in the path from the root to the $t$-th leaf node. 
Return the private sum by the following procedure: 
(a) Initialize $s=0$ and convert $t$ into binary representation $[r_1,\cdots, r_h]$ with $r_h$ being the smallest bit. 
(b) For each $j\in[h]$, if $r_j=1$, add the value of the left sibling of node $u_j$ to $s$. If $u_j$ is the left sibling, add itself to $s$.

\section{Proofs for Theorems in Section~\ref{sec:meta_alg}}\label{sec:proof_meta}
\subsection{Proof of Theorem~\ref{thm:DP_meta}}\label{sec:proof_meta_privacy}
\begin{proof}
We can denote the primal update in Line 2-4 of Algorithm~\ref{alg:algorithm_meta} as a function $x_t=\pi_t(y_t,\mu_t,\tilde{B}_{t})$. By the condition in Theorem~\ref{thm:DP_meta}, the mechanism to privatize the dual variables and remaining budgets $\{(\mu_t,\tilde{B}_{t}), t\in [T]\}$ are  $(\epsilon,\delta)-$DP with respect to all the inputs $\{y_t,t\in [T]\}$. Thus, the conditions of Billboard Lemma~\ref{lma:DP_JDP} are satisfied. Thus,  the composition of $\{x_1,\cdots,x_T\}$ preserves $(\epsilon,\delta)-$JDP.

Next, we prove that \ouralg satisfies $(\epsilon,\delta)-$LDP.  According to the definition of LDP (Definition~\ref{def:LDP}), $(\epsilon, \delta)-$LDP holds if a local mechanism $\mathcal{M}(y)$, which maps local context to the information communicated with other users, satisfies the $(\epsilon, \delta)-$DP. In Algorithm~\ref{alg:algorithm_meta}, the local mechanisms are $\mathcal{M}_{B}$ and $\mathcal{M}_{D}$ which output the privatized remaining budget $\tilde{B}_{t+1}$ and the dual variable $\mu_{t+1}$, respectively.  Thus, if the composition of the remaining budgets $(\tilde{B}(1),\cdots, \tilde{B}(T))$ and dual variables $(\mu_{1},\cdots,\mu_{T})$ satisfies $(\epsilon, \delta)-$DP, Algorithm 1 satisfies $(\epsilon, \delta)-$LDP.  

\end{proof}

\subsection{Proof of Theorem~\ref{thm:reward_meta}}\label{sec:proof_meta_reward}
\subsubsection{Proof of constraint feasibility}
By Condition (1) in Theorem~\ref{thm:reward_meta}, we have $B_{t}\geq \tilde{B}_{t}- B_{\mathrm{gap},t}$.
Thus, at each round $t$, by ensuring $b_{t}(x_{t})\leq \tilde{B}_{t}- B_{\mathrm{gap},t}$ with probability at least $1-\zeta$, Algorithm \ref{alg:algorithm_meta} always guarantees the constraint $\sum_{t=1}^{T}b_t(x_t)\leq B_1$ with probability at least $1-\zeta$ for any horizon $T$. 
\subsubsection{Proof for Adversarial Setting}
\begin{proof}
Let $\tau_A$ be the first time slot when there exists a $j\in [m]$ such that $\bar{b}_j\geq \tilde{B}_{\tau_A+1,j}-B_{\mathrm{gap},\tau_A+1}$, which implies that $\tau_A$, typically termed as "stopping time", is a random variable depending on the randomness of the algorithm. 

If $t\leq \tau_A$, since $x_t=\arg\max_{x\in \mathcal{X}}\{f_t(x)-\mu_t^\top b_t(x)\}$, we have for the optimal action $x_t^*$, $f_t(x_t^*)-\mu_t^\top b_t(x_t^*)\leq f_t(x_t)-\mu_t^\top b_t(x_t)$. Also, we have $0\leq f_t(x_t)-\mu_t^\top b_t(x_t)$ because $f_t(0)=0$ and $b_t(0)=0$. 
Combining them, we also have the following inequality held.
\begin{equation}\label{eqn:metaproof_1}
\begin{split}
\alpha f_t(x_t)&=f_t(x_t)+(\alpha-1)f_t(x_t)\\
&\geq f_t(x_t)-\mu_t^\top b_t(x_t)+\mu_t^\top b_t(x_t)+(\alpha-1)f_t(x_t)\\
&\geq f_t(x_t^*)-\mu_t^\top b_t(x_t^*)+\mu_t^\top b_t(x_t)+(\alpha-1)\mu_t^\top b_t(x_t)\\
&= f_t(x_t^*)-\alpha \mu_t^{\top}(\rho-b_t(x_t))- \mu_t^\top b_t(x_t^*)+\alpha \mu_t^\top \rho\\
&\geq f_t(x_t^*)-\alpha w_t(\mu_t),
\end{split}
\end{equation}
where the last inequality holds because 
$\alpha=\sup_{j\in[m]}\frac{\bar{b}_j}{\rho_j}$ and so $\alpha \mu_t^\top \rho\geq \mu_t^\top \bar{\bm{b}}\geq \mu_t^\top b_t(x_t^*)$ where $\bar{\bm{b}}=[\bar{b}_1,\cdots, \bar{b}_m]$.

Given a stopping time $\tau_A$, with probability at least $1-\zeta$, $\zeta\in(0,1)$, we can write the reward difference as
\begin{equation}\label{eqn:meta_proof2}
\begin{split}
&\sum_{t=1}^Tf_t(x_t^*)-\alpha \sum_{t=1}^Tf_t(x_t)\\
\leq &\sum_{t=1}^Tf_t(x_t^*)-\alpha \sum_{t=1}^{\tau_A}f_t(x_t)\\
\leq &\sum_{t=1}^Tf_t(x_t^*)-\sum_{t=1}^{\tau_A}\left(f_t(x_t^*) -\alpha w_t(\mu_t)\right)\\
\leq & (T-\tau_A)\bar{f}+\alpha \sum_{t=1}^{\tau_A}w_t(\mu)+\alpha \Gamma(T,\zeta)
\end{split}
\end{equation}
where the second inequality holds by \eqref{eqn:metaproof_1}, the third inequality holds with probability at least $1-\zeta$ over the randomness of the algorithm for any $\mu\in\mathcal D$ by Condition (2) in Theorem~\ref{thm:reward_meta}.

If $\tau_A=T$, set $\mu=0$. Then, we have with probability at least $1-\zeta$,
\begin{equation}
 \sum_{t=1}^Tf_t(x_t^*)-\alpha \sum_{t=1}^Tf_t(x_t)
\leq  \alpha \Gamma(T,\zeta)
\end{equation}

Otherwise, there exists a resource $j$ such that 
\(\bar{b}_j\geq \tilde{B}_{\tau_A+1,j}-B_{\mathrm{gap},\tau_A+1}\). By the condition that $\forall t\in [T], \|\tilde{B}_t-B_t\|\leq B_{\mathrm{gap},t}\leq \Gamma(T,\zeta)$, it holds that 
\begin{equation}\label{eqn:meta_proof3}
\sum_{t=1}^{\tau_A}b_{t,j}(x_{t})= B_j-B_{\tau_A+1,j} \geq B_j-\tilde{B}_{\tau_A+1,j}-B_{\mathrm{gap},\tau_A+1}\geq  T\rho_j-\bar{b}_j-2\Gamma(T,\zeta)
\end{equation}

Next we choose $\mu=\mu_j\mathbf{e}_j$ and we can bound the complementary slackness with probability at least $1-\zeta$ as
\begin{equation}\label{eqn:meta_proof4}
\begin{split}
&\alpha\sum_{t=1}^{\tau_A}w_t(\mu)=\alpha\sum_{t=1}^{\tau_A}\mu^\top (\rho -b_t(x_t))\\
&=\alpha\sum_{t=1}^{\tau_A}\mu_j (\rho_j -b_{t,j}(x_{t}))\\
&= \alpha \tau_A\mu_j\rho_j -\alpha \mu_j\sum_{t=1}^{\tau_A} b_{t,j}(x_{t})\\
&\leq -\alpha (T-\tau_A)\mu_j\rho_j  +\alpha \mu_j\bar{b}_j+2\alpha\mu_j \Gamma(T,\zeta),
\end{split}
\end{equation}
where the inequality holds by \eqref{eqn:meta_proof3}.

By choosing $\mu_j=\frac{\bar{f}}{\alpha \rho_j}$, we have
\begin{equation}\label{eqn:meta_proof5}
\begin{split}
\alpha\sum_{t=1}^{\tau_A}w_t(\mu)
\leq -(T-\tau_A)\bar{f}+ \frac{\bar{b}_j}{\rho_j}\bar{f}+\frac{2\bar{f}}{\rho_j} \Gamma(T,\zeta)
\end{split}
\end{equation}
Substituting the bound of $\alpha\sum_{t=1}^{\tau_A}w_t(\mu)$ back to \eqref{eqn:meta_proof2}, with probability of at least $1-\zeta, \zeta\in(0,1)$ over the randomness of the algorithm, it holds for any stopping time $\tau_A$ (either $\tau_A<T$ or $\tau_A=T$) that
\begin{equation}\label{eqn:violation_condition}
\begin{split}
\sum_{t=1}^T&f_t(x_t^*)-\alpha \sum_{t=1}^Tf_t(x_t)
\leq \alpha \bar{f}+(\frac{2\bar{f}}{\hat{\rho}}+\alpha) \Gamma(T,\zeta),
\end{split}
\end{equation}
where $\hat{\rho}=\min_{j\in[m]}\rho_j$.

Since the algorithm uses $\zeta=\zeta_T\in (0,1)$ satisfying \(\lim_{T\to\infty}\zeta_T=0, \lim_{T\to\infty}\frac{\Gamma(T,\zeta_T)}{T}=0\), as $T\rightarrow \infty$, for any fixed sequence $\bm{y}\in\mathcal{Y}^T$, we have with probability of at least $(1-\zeta_T)\rightarrow 1$ over the randomness of the algorithm that 
\begin{equation}
\begin{split}
\frac{1}{T}(\sum_{t=1}^T&f_t(x_t^*)-\alpha \sum_{t=1}^Tf_t(x_t))
\leq \frac{\alpha \bar{f}}{T}+\frac{(\frac{2\bar{f}}{\hat{\rho}}+\alpha)\Gamma(T,\zeta_T)}{T}\rightarrow 0,
\end{split}
\end{equation}
This establishes an asymptotic competitive ratio of at least $1/\alpha$ in probability, as defined in Definition~\ref{def:cr}.
\end{proof}

\subsubsection{Proof for i.i.d. Stochastic Setting}
\begin{proof}
All expectations below are over the i.i.d.\ contexts and the
algorithm's internal randomness, unless stated otherwise.
Define the expected dual objective as
\[
\bar D(\mu\mid\mathcal P)
=\frac{1}{T}\mathbb E[D(\mu\mid\bm y)]
=\mathbb E[f^*(\mu)]+\mu^\top\rho,
\qquad
f_t^*(\mu)
=\max_{x\in\mathcal X_t}\{f_t(x)-\mu^\top b_t(x)\}.
\]
Let $\xi_t=\{(y_1,v_1),\ldots,(y_t,v_t)\}$, where $v_t$
collects the internal randomness used by the algorithm at round $t$,
and let $\varsigma(\xi_t)$ denote the corresponding history.
Since $\mu_t$ is $\varsigma(\xi_{t-1})$-measurable and $y_t$
is independent of this history, it holds that
\(
\mathbb E[f_t^*(\mu_t)\mid\varsigma(\xi_{t-1})]
=\bar D(\mu_t\mid\mathcal P)-\mu_t^\top\rho.
\)
The null action, nonnegative costs, and nonnegative dual variables
ensure that $0\leq f_t^*(\mu_t)\leq\bar f$.

Define the martingale
\[
Z_t=\sum_{s=1}^{t}
\left(
\mathbb E[f_s^*(\mu_s)\mid\varsigma(\xi_{s-1})]
-f_s^*(\mu_s)
\right), Z_0=0.
\]
Its increments are bounded in absolute value by $\bar f$.
Use the stopping time $\tau_A\in\{0,\ldots,T\}$ from the
adversarial proof: it is the first $t<T$ for which some resource
$j$ satisfies
$\tilde B_{t+1,j}-B_{\mathrm{gap},t+1}\leq\bar b_j$,
and is $T$ if no such time exists.
Thus, $x_t=\hat x_t$ for every $t\leq\tau_A$.
Since $\tau_A\leq T$, the optional stopping theorem gives
\(
\mathbb E[Z_{\tau_A}]=0,
|Z_{\tau_A}|\leq T\bar f.
\)
For $t\leq\tau_A$, the primal update satisfies
\begin{equation}
\begin{aligned}
f_t(x_t)
&=f_t^*(\mu_t)+\mu_t^\top b_t(x_t)\\
&= \mathbb E[f_t^*(\mu_t)\mid\varsigma(\xi_{t-1})]+\mu_t^\top\rho+\mu_t^\top (b_t(x_t)-\rho)+ f_t^*(\mu_t)-\mathbb E[f_t^*(\mu_t)\mid\varsigma(\xi_{t-1})]\\
&=\bar D(\mu_t\mid\mathcal P)-w_t(\mu_t)
  -(Z_t-Z_{t-1}).
\end{aligned}
\end{equation}

By weak duality, for every fixed $\mu\in\mathbb R_+^m$, we have
\[
\mathbb E[\mathrm{OPT}(\bm y)]
\leq T\bar D(\mu\mid\mathcal P).
\]
Because this inequality holds for every $\mu$, the deterministic
quantity $\mathbb E[\mathrm{OPT}(\bm y)]/T$ also lower bounds
$\bar D(\mu_t\mid\mathcal P)$ for every realized $\mu_t$.
Using nonnegative rewards after $\tau_A$ and
$\mathbb E[\mathrm{OPT}(\bm y)]\leq T\bar f$, we obtain,
for each realization,
\begin{equation}\label{eq:meta_iid_pathwise}
\begin{aligned}
&\mathbb E[\mathrm{OPT}(\bm y)]-F_T^{\mathcal A}(\bm y)\\
\leq &
\mathbb E[\mathrm{OPT}(\bm y)]
-\sum_{t=1}^{\tau_A}\bar D(\mu_t\mid\mathcal P)
+\sum_{t=1}^{\tau_A}w_t(\mu_t)+Z_{\tau_A}\\
\leq&\mathbb E[\mathrm{OPT}(\bm y)]
-\tau_A\bar D(\bar{\mu}_A\mid\mathcal P)
+\sum_{t=1}^{\tau_A}w_t(\mu_t)+Z_{\tau_A}\\
\leq&
\left(1-\frac{\tau_A}{T}\right)
\mathbb E[\mathrm{OPT}(\bm y)]
+\sum_{t=1}^{\tau_A}w_t(\mu_t)+Z_{\tau_A}\\
\leq&
(T-\tau_A)\bar f
+\sum_{t=1}^{\tau_A}w_t(\mu_t)+Z_{\tau_A},
\end{aligned}
\end{equation}
where $\bar{\mu}_A=\frac{1}{\tau_A}\sum_{t=1}^{\tau_A}\mu_t$, the second inequality holds because the dual objective $\bar{D}(\bar{\mu}_A\mid \mathcal{P})$ is convex, the third inequality holds by weak duality, and the last inequality holds .

Let $\mathcal E$ be the event on which Condition~(1) and 
Condition~(2) of Theorem~\ref{thm:reward_meta} both hold.
The union bound gives $\Pr(\mathcal E^c)\leq2\zeta$:
the bound holds for every fixed context sequence and therefore
also after averaging over the i.i.d.\ contexts.
On $\mathcal E$, Condition~(2) and
\eqref{eq:meta_iid_pathwise} imply that for every
$\mu\in\mathcal D$,
\begin{equation}\label{eq:meta_iid_good_event}
\begin{aligned}
\mathbb E[\mathrm{OPT}(\bm y)]-F_T^{\mathcal A}(\bm y)\leq
(T-\tau_A)\bar f+\sum_{t=1}^{\tau_A}w_t(\mu)
+\Gamma(T,\zeta)+Z_{\tau_A}.
\end{aligned}
\end{equation}

If $\tau_A=T$, choose $\mu=\bm0$.
Otherwise, there exists a resource $j$ that satisfies \(\bar{b}_j\geq \tilde{B}_{\tau_A+1,j}-B_{\mathrm{gap},\tau_A+1}\). Choose such a resource $j$ and set $\mu=(\bar f/\rho_j)\mathbf e_j\in\mathcal D$. 
On $\mathcal E$, we have by Condition~(1) and (2) that
\(
B_{\tau_A+1,j}
\leq\tilde B_{\tau_A+1,j}+B_{\mathrm{gap},\tau_A+1}
\leq\bar b_j+2\Gamma(T,\zeta),
\)
and hence it holds that
\[
\sum_{t=1}^{\tau_A}b_{t,j}(x_t)
=T\rho_j-B_{\tau_A+1,j}
\geq T\rho_j-\bar b_j-2\Gamma(T,\zeta).
\]
It follows that
\begin{equation}
\begin{split}
\sum_{t=1}^{\tau_A}w_t(\mu)&=\sum_{t=1}^{\tau_A}\mu^\top (\rho -b_t(x_t))\\
&= \tau_A\mu_j\rho_j -\mu_j\sum_{t=1}^{\tau_A} b_{t,j}(x_{t})\\
&\leq -(T-\tau_A)\bar f
+\frac{\bar b_j}{\rho_j}\bar f
+\frac{2\bar f}{\rho_j}\Gamma(T,\zeta),
\end{split}
\end{equation}
where the inequality holds by the bound of \(\sum_{t=1}^{\tau_A}b_{t,j}(x_t)\) and the choice of $\mu$.

Combining both cases of $\tau_A$ in \eqref{eq:meta_iid_good_event}, we have on $\mathcal E$, 
\begin{equation}
\begin{aligned}
\mathbb E[\mathrm{OPT}(\bm y)]-F_T^{\mathcal A}(\bm y)\leq
\alpha\bar f
+\left(\frac{2\bar f}{\hat\rho}+1\right)\Gamma(T,\zeta)
+Z_{\tau_A},
\end{aligned}
\end{equation}
where $\alpha=\max_{j\in[m]}\bar b_j/\rho_j$ and
$\hat\rho=\min_{j\in[m]}\rho_j>0$.

On $\mathcal E^c$, the difference between expected reward and the algorithm reward is unconditionally upper bounded due to the nonnegative rewards:
$\mathbb E[\mathrm{OPT}(\bm y)]-F_T^{\mathcal A}(\bm y)
\leq T\bar f$.
Taking expectations over the event $\mathcal E$ and $\mathcal E^c$, we have
\begin{equation}\label{eq:meta_iid_expected_regret}
\begin{aligned}
&\mathbb E\left[
\mathrm{OPT}(\bm y)-F_T^{\mathcal A}(\bm y)
\right]\\
\leq &
\alpha\bar f
+\left(\frac{2\bar f}{\hat\rho}+1\right)\Gamma(T,\zeta)
+\mathbb E[Z_{\tau_A}\mathbf1_{\mathcal E}]
+T\bar f\,\Pr(\mathcal E^c)\\
=&
\alpha\bar f
+\left(\frac{2\bar f}{\hat\rho}+1\right)\Gamma(T,\zeta)
-\mathbb E[Z_{\tau_A}\mathbf1_{\mathcal E^c}]
+T\bar f\,\Pr(\mathcal E^c)\\
\leq &
\alpha\bar f
+\left(\frac{2\bar f}{\hat\rho}+1\right)\Gamma(T,\zeta)
+2T\bar f\zeta,
\end{aligned}
\end{equation}
where the first equality holds because  $\mathbb E[Z_{\tau_A}]
=\mathbb E[Z_{\tau_A}\mathbf1_{\mathcal E}]
+\mathbb E[Z_{\tau_A}\mathbf1_{\mathcal E^c}]=0$, and the last
inequality holds by $|Z_{\tau_A}|\leq T\bar f$.

Finally, configure the algorithm with $\zeta=\zeta_T$ as in
Condition~(2). Since $\zeta_T\to0$, \(T\zeta_T=O(1)\) and
$\Gamma(T,\zeta_T)/T\to0$, it follows by \eqref{eq:meta_iid_expected_regret}
that
\[
\mathbb E\left[
\mathrm{OPT}(\bm y)-F_T^{\mathcal A}(\bm y)
\right] \leq \alpha\bar f
+\left(\frac{2\bar f}{\hat\rho}+1\right)\Gamma(T,\zeta)
+2T\bar f\zeta_T =o(T),
\]
thus establishing the sublinear expected regret.
\end{proof}

\subsubsection{Discussion on the Extension to the Non-stationary Setting}\label{sec:non-stationary_discussions}

A more general stochastic input model is the non-stationary setting, in which the context distribution varies over time, i.e. the input distributions in
\(
\mathcal{P}_{1:T}=\{\mathcal{P}_1,\ldots,\mathcal{P}_T\}
\)
are not necessarily identical. One way to quantify the degree of non-stationarity is through the mean total variance deviation
\begin{equation}\label{eqn:totalvariance}
\mathrm{TV}(\mathcal{P}_{1:T})
=
\sum_{t=1}^T
\mathbb{D}_{\mathrm{TV}}\!\left(
\mathcal{P}_t,\,
\bar{\mathcal{P}}
\right),
\qquad
\bar{\mathcal{P}}
:=
\frac{1}{T}\sum_{t=1}^T \mathcal{P}_t,
\end{equation}
where \(\mathbb{D}_{\mathrm{TV}}\) denotes the total variation distance between two distributions. A larger value of \(\mathrm{TV}(\mathcal{P}_{1:T})\) indicates a more rapidly changing context distribution, making it harder for an online algorithm to track the distribution shift.

Without additional assumptions, \cite{dual_mirror_descent_online_allocation_balseiro2020dual} show that online allocation algorithms can incur linear expected regret in the non-stationary setting even in the absence of privacy constraints. Therefore, under the considered DP notions, one may not expect sublinear regret either.

Nevertheless, the analysis for the i.i.d.\ stochastic setting provides a useful starting point for extending performance bounds to non-stationary stochastic settings by adapting arguments similar to those in Theorem 3 of \cite{dual_mirror_descent_online_allocation_balseiro2020dual}. In particular, their analysis shows that non-stationarity contributes an additional term in the expected regret that scales with \(\mathrm{TV}(\mathcal{P}_{1:T})\).

\section{Proofs for Theorems in Section~\ref{ppoa_dmd_algorithm}}\label{sec:proof_ppoa_dmd}

\subsection{Proof of Theorem \ref{thm:privacy_local}}\label{sec:proof_JDP_local}
\begin{proof}
The Global Sensitivity of the consumption function is bounded as
\begin{equation}
  \Delta_b = \sup_{ \mathcal{D}_t,\mathcal{D}_t'}\left\|b_t(\mathcal{D}_t)-b'_t(\mathcal{D}_t')\right\|_2 =\sup_{ b_t,b_t',x_t,x_t'}\left\|b_t(x_t)-b_t'(x_t')\right\|_2\leq 2\bar{b},
\end{equation}
where the inequality holds by triangle inequality. 
Thus, by Gaussian mechanism for RDP in Lemma~\ref{Gaussian_Mechnism_RDP}, the consumption mechanism $\tilde{b}_t(x_t)=b_t(x_t)+n_t$ satisfies $(\alpha, \frac{2\alpha \bar{b}^2}{\sigma^2})$-RDP. By Lemma \ref{lma:RDP_composition}, the composition $(\tilde{b}_1,\cdots, \tilde{b}_{t+1})$ satisfies $(\alpha, \frac{2\alpha \bar{b}^2}{\sigma^2})$-RDP.

Since the budget dynamic and the dual mirror descent are all post processing, the composition of dual variables $(\mu_1,\cdots, \mu_T)$ and remaining budgets $(\tilde{B}_1,\cdots, \tilde{B}_T)$ both satisfy $(\alpha, \frac{2\alpha \bar{b}^2}{\sigma^2})$-RDP. 
By Lemma \ref{lma:DP-RDP}, the composition of variables $((\mu_1,\tilde{B}_1),\cdots, (\mu_T,\tilde{B}_T))$  satisfies $(\frac{2\alpha \bar{b}^2}{\sigma^2}+\frac{\ln(1/\delta)}{\alpha-1},\delta)$-DP with $\delta\in (0,1)$ and $\alpha>1$.
Optimally choosing $\alpha=1+\sqrt{\frac{\ln(1/\delta)\sigma^2}{2\bar{b}^2}}$, the composition  satisfies $(\frac{2\bar{b}^2}{\sigma^2}+2\sqrt{\frac{2\bar{b}^2}{\sigma^2}\ln(1/\delta)},\delta)$-DP. Choosing $\sigma=\frac{\sqrt{2}\bar{b}(\sqrt{\ln(1/\delta)}+\sqrt{\ln(1/\delta)+\epsilon})}{\epsilon}$, we can guarantee $(\epsilon,\delta)-$DP, which proves the condition in Theorem~\ref{thm:DP_meta}.
\end{proof}

\subsection{Proof of Theorem~\ref{thm:reward_local}}\label{sec:proof_constraint_local}
We prove the two conditions in Theorem~\ref{thm:reward_meta} are satisfied by \ouralgdmd.
\subsubsection{Proof of Condition (1)}
\begin{proof}
  The true remaining budget at round $t$ can be expressed as $B_t=B_1-\sum_{s=1}^{t-1}b_{s}(x_{s})$ and the corrupted remaining budget is expressed as $\tilde{B}_{t}=B_1-\sum_{s=1}^{t-1}\tilde{b}_{s}(x_{s})=B_{1}-\sum_{s=1}^{t-1}(b_{s}(x_{s})+n_{s})$. 

  Let $v_t=\sum_{i=1}^t n_{i}$. Observing that $\{n_{s}\}_{s\in[t]}\in\mathcal{N}(0,\sigma^2)$ are independent Gaussian random variables, we have $v_t\sim \mathcal{N}(0,t\sigma^2\bm{I})$.   By Lemma~\ref{lma:Gaussian_concentration} and union bounds, given a resource $j\in [m]$, it holds $\forall t\in [T]$ with probability at least $1-\zeta $ that, 
\begin{equation}
|v_{t,j}|\leq \sigma \sqrt{2t\ln\frac{2 T}{\zeta}},
\end{equation}
 By union bound, it holds  with probability at least $1-\zeta$ that  
 \begin{equation}\label{eqn:concentration_norm_inf_local}
   \| B_t-\tilde{B}_t\|_{\infty}=\| v_t\|_{\infty} \leq  \sigma \sqrt{2t\ln\frac{2 mT}{\zeta}}=B_{\mathrm{gap},t}.  
 \end{equation}
 
\end{proof}

\subsubsection{Proof of Condition (2)}
To prove Condition (2), we first prove the convergence of projected mirror descent in Lemma~\ref{lma:dual_convergence} and the concentration of $\sum_{t=1}^T\mu_t^\top n_t$ given a norm-bounded $\mu_t$ in Lemma~\ref{lma:weighted_sum_gaussian}.

\begin{lemma}\label{lma:dual_convergence}
If $\mu_t$ is updated as $\mu_{t+1}=\arg\min_{\mu:\|\mu\|_2\leq \mu_{\max}} \tilde{g}_t^\top\mu+\frac{1}{\eta}V_h(\mu,\mu_t)$ where $\tilde{g}_t=g_t+n_t,n_t\sim\mathcal{N}(0,\sigma\bm{I})$ and $\|g_t\|_2\leq G$ for $T$ rounds, then for any $\mu: \|\mu\|_2\leq \mu_{\max}$, it holds for $\tilde{w}_t(\mu_t)=\tilde{g}_t^\top \mu_t$ with probability at least $1-\zeta,\zeta\in (0,1)$ that
\begin{equation}
\sum_{t=1}^T \tilde{w}_t(\mu_t)-\tilde{w}_t(\mu)\leq \eta \left(G^2+m\sigma^2\right)T+\frac{1}{\eta}V_h(\mu,\mu_1)+\eta\sigma^2\sqrt{mT\ln(1/\zeta)}+\eta\sigma^2\ln(1/\zeta).
\end{equation}
\end{lemma}
\begin{proof}

Since $\sum_{t=1}^T\|n_t/\sigma\|^2$ is a $\chi^2$-distribution with $mT$ degrees of freedom, by Laurent-Massart bound (Lemma ~\ref{lma:LM_bound}), we have $\Pr(\sum_{t=1}^T\|n_t/\sigma\|^2\geq mT+2\sqrt{m Tx}+2x)\leq \exp(-x)$ and further implies with probability at least $1-\zeta$, we have $\|n_t\|^2\leq mT\sigma^2+2\sigma^2\sqrt{mT\ln(1/\zeta)}+2\sigma^2\ln(1/\zeta)$. 
Thus, It holds with probability at least $1-\zeta$ for the perturbed sug-gradient that 
\begin{equation}
\begin{split}   \sum_{t=1}^T\|\tilde{g}_t\|_2^2&=\sum_{t=1}^T\|g_t+n_t\|_2^2\leq 2\sum_{t=1}^T\left(\|g_t\|_2^2+\|n_t\|_2^2\right)\\
  &\leq   2 T\left(G^2+m\sigma^2\right)+4\sigma^2\sqrt{mT\ln(1/\zeta)}+2\sigma^2\ln(1/\zeta)
\end{split}
\end{equation}
Replace $g_t$ in Lemma~\ref{lma:DMD} with $\tilde{g}_t$,
it holds with probability at least $1-\zeta$ that
\begin{equation}
\begin{split}
&\sum_{t=1}^T \tilde{w}_t(\mu_t)-\tilde{w}_t(\mu)\\
\leq &\eta T\left(G^2+m\sigma^2\right)+\frac{1}{\eta}V_h(\mu,\mu_1)+2\eta\sigma^2\sqrt{mT\ln(1/\zeta)}+\eta\sigma^2\ln(1/\zeta)
\end{split}
\end{equation}
\end{proof}

\begin{lemma}\label{lma:weighted_sum_gaussian}
If $\forall t\in [T]$, $\|\mu_t\|_2\leq \mu_{\max}$. It holds with probability at least $1-\zeta, \zeta\in (0,1)$ that 
    \begin{equation}
    \sum_{t=1}^{T}\mu_t^\top n_t \geq  - \mu_{\max} \sigma\sqrt{2T \ln(1/\zeta)}
\end{equation}
\end{lemma}

\begin{proof}
 Define a sequence $\xi_t=\{(y_1,n_1),\cdots, (y_t,n_t)\}$ and let $\mathcal{F}_t$ be the sigma-algebra generated by $\xi_t$, then $\{\mathcal{F}_t\}_{t\in [T]}$ is a Filtration. Therefore, $\mu_t$ is $\mathcal{F}_{t-1}$-measurable and $n_t \sim \mathcal{N}(0,\sigma^2 I)$ are independent across $t$. 
Then we have 
\[
\mathbb{E}[\mu_t^\top n_t\mid \mathcal{F}_{t-1}]
= \mu_t^\top \mathbb{E}[n_t \mid \mathcal{F}_{t-1}]
= 0.
\]
Thus, conditional on $\mathcal{F}_{t-1}$, $\mu_t^\top n_t$ is Gaussian with mean $0$ and variance
\[
\operatorname{Var}(\mu_t^\top n_t \mid \mathcal{F}_{t-1})
= \sigma^2 \|\mu_t\|_2^2
\le \sigma^2 \mu_{\max}^2.
\]
Hence, by Gaussian Moment Generation function, for every $\lambda \in \mathbb{R}$,
\[
\mathbb{E}\!\left[e^{\lambda \mu_t^\top n_t} \mid \mathcal{F}_{t-1}\right]
= \exp\!\left(\frac{\lambda^2 \sigma^2 \|\mu_t\|_2^2}{2}\right)
\le \exp\!\left(\frac{\lambda^2 \sigma^2 \mu_{\max}^2}{2}\right).
\]
Let $W_t=\sum_{s=1}^t \mu_s^\top n_s$. It holds for any $t\in [T]$ that
\begin{equation}\label{eqn:power_property}
\mathbb{E}\!\left[e^{\lambda W_t}\right] = \mathbb{E}\!\left[e^{\lambda W_{t-1}}\mathbb{E}\!\left[e^{\lambda \mu_t^\top n_t}\mid \mathcal{F}_{t-1}\right]\right]\leq \exp\!\left(\frac{\lambda^2 \sigma^2 \mu_{\max}^2}{2}\right) \mathbb{E}\!\left[e^{\lambda W_{t-1}}\right]
\end{equation}
Applying \eqref{eqn:power_property} iteratively given $W_0=0$, we have
\[
\mathbb{E}\!\left[e^{\lambda W_T}\right]
\le \exp\!\left(\frac{\lambda^2 T \sigma^2 \mu_{\max}^2}{2}\right).
\]
Therefore, for any $u > 0$ and any $\lambda > 0$, Chernoff's bound yields
\[
\Pr(W_T \le -u)
\le e^{-\lambda u} \mathbb{E}\!\left[e^{-\lambda W_T}\right]
\le \exp\!\left(-\lambda u + \frac{\lambda^2 T \sigma^2 \mu_{\max}^2}{2}\right).
\]
Optimizing over $\lambda$ at
\(
\lambda = \frac{u}{T \sigma^2 \mu_{\max}^2}
\)
gives
\[
\Pr(W_T \le -u)
\le \exp\!\left(-\frac{u^2}{2T \sigma^2 \mu_{\max}^2}\right).
\]
Taking
\(
u = \sigma \mu_{\max} \sqrt{2T \ln(1/\zeta)}
\)
shows that, with probability at least $1-\zeta$,
\[
\sum_{t=1}^T \mu_t^\top n_t
\ge -\sigma \mu_{\max} \sqrt{2T \ln(1/\zeta)}.
\]
\end{proof}

\textbf{Proof of Theorem~\ref{thm:reward_local}}
\begin{proof}
    We are ready to prove the convergence bound of the projected private DMD in Algorithm~\ref{alg:algorithm_local}.  Denote the privatized dual objective as $\tilde{w}_t(\mu)=\tilde{g}_t^\top \mu$ where $\tilde{g}_t=\rho-b_t(x_t)+n_t$ and $n_t\sim\mathcal{N}(0,\sigma^2\mathbf{I})$. 
    Let $\mu_1,\cdots,\mu_T$ be the dual variables given by the projected private DMD.  The clean dual objective can be expressed as
\begin{equation}
\begin{split}
   &\sum_{t=1}^T[w_t(\mu_t)-w_t(\mu)]\\
   =&\sum_{t=1}^T [\tilde{w}_t(\mu_t)-\tilde{w}_t(\mu)]-\sum_{t=1}^T n_t\mu_t + \sum_{t=1}^Tn_t\mu\\
   \leq& \eta \left(G^2+m\sigma^2\right)T+\frac{1}{\eta}V_h(\mu,\mu_1)+2\eta\sigma^2\sqrt{mT\ln(3/\zeta)}+\eta\sigma^2\ln(3/\zeta)\\
   &+\mu_{\max} \sigma\sqrt{2T \ln(3/\zeta)}+\mu \sigma \sqrt{2T\ln\frac{6 mT}{\zeta}}
\end{split}
\end{equation}
where the inequality holds by Lemma~\ref{lma:dual_convergence}, Lemma~\ref{lma:weighted_sum_gaussian}, and \eqref{eqn:concentration_norm_inf_local} and union bounds. Choosing $\eta=\sqrt{\frac{ V_h(\mu,\mu_1)}{T(G^2+m\sigma^2)}}$, it holds that
\begin{equation}
\begin{split}
   &\sum_{t=1}^T[w_t(\mu_t)-w_t(\mu)]\\
   \leq& 2\sqrt{V_h(\mu,\mu_1)\left(G^2+m\sigma^2\right)T}+\mu_{\max} \sigma\sqrt{2T \ln(3/\zeta)}\\
   &+\mu \sigma \sqrt{2T\ln\frac{6 mT}{\zeta}} +\sqrt{V_h(\mu,\mu_1)}\left(2\sigma\sqrt{m\ln(3/\zeta)}+\sigma\frac{\ln(3/\zeta)}{\sqrt{T}}\right)\\
   =&\mathcal{O}(\sqrt{T\ln(T/\zeta)}\sigma\mu_{\max})= \mathcal{O}(\sqrt{T\ln(T/\zeta)}\mu_{\max}\frac{\sqrt{\log(1/\delta)}}{\epsilon}),
\end{split}
\end{equation}
where the last equality holds by the choice of $\sigma=\frac{\sqrt{2}\bar{b}(\sqrt{\ln(1/\delta)}+\sqrt{\ln(1/\delta)+\epsilon})}{\epsilon}$.

We define $\Gamma(T,\zeta)=\mathcal{O}(\sqrt{T\ln(T/\zeta)}\mu_{\max}\frac{\sqrt{\log(1/\delta)}}{\epsilon})$
By \eqref{eqn:concentration_norm_inf_local} in the proof of condition (1), we have \[B_{\mathrm{gap},T}=\sigma \sqrt{2T\ln\frac{2 mT}{\zeta}}\leq\mathcal{O}(\sqrt{T\ln(T/\zeta)}\mu_{\max}\frac{\sqrt{\log(1/\delta)}}{\epsilon}),\] thus satisfying the requirement of Condition~(2) on $B_{\mathrm{gap},T}$.

Moreover, by choosing $\zeta=\zeta_T=\frac{1}{T^{\vartheta}+1}$ given a constant $\vartheta\geq 1$, we have \[\Gamma(T,\zeta_T)=\mathcal{O}(\sqrt{T\ln(T)}\mu_{\max}\frac{\sqrt{\log(1/\delta)}}{\epsilon})=\widetilde{\mathcal{O}}(\sqrt{T}\mu_{\max}\frac{\sqrt{\log(1/\delta)}}{\epsilon}).\]
Therefore, conditions in (2) of Theorem~\ref{thm:reward_meta} are satisfied. 
    
\end{proof}

\section{Proofs for Theorems in Section~\ref{sec:central_Algorithm}}\label{sec:proof_PPOA_ftrl}

\subsection{Proof of Theorem \ref{thm:privacy_ftrl}}\label{sec:proof_JDP_central}
\begin{proof}
By Billboard Lemma (Lemma~\ref{lma:DP_JDP}), if $(s_1,\cdots, s_T)$ satisfies $(\epsilon, \delta)-$ differential privacy, then the composition $(\mu_1,\cdots, \mu_T)$ satisfies $(\epsilon, \delta)-$ differential privacy, and the composition $(x_1,\cdots, x_T)$ satisfies $(\epsilon, \delta)-$JDP. Thus, we only need to prove  the condition that $(s_1,\cdots, s_T)$ satisfies $(\epsilon, \delta)-$ differential privacy. 

The Global Sensitivity of the consumption function is bounded as
\begin{equation}
  \Delta_b = \sup_{ \mathcal{D}_t,\mathcal{D}_t'}\left\|b_t(\mathcal{D}_t)-b'_t(\mathcal{D}_t')\right\|_2 =\sup_{ b_t,b_t',x_t,x_t'}\left\|b_t(x_t)-b_t'(x_t')\right\|_2\leq 2\bar{b},
\end{equation}
where the inequality holds by triangle inequality.
Thus, by the tree initialization procedure in Algorithm \ref{alg:algorithm_ftrl}, each node in the tree satisfies $(\alpha, \frac{2\alpha \bar{b}^2}{\sigma^2})-$RDP by Lemma~\ref{Gaussian_Mechnism_RDP}. Since each leaf node affects at most $\left\lceil \log_2(T)\right\rceil$ nodes which locate at the path from the leaf node to the root of the tree, by Lemma \ref{lma:RDP_dependent_composition} (Proposition 1 in \cite{Renyi_DP_mironov2017renyi}),  the composition of all the nodes in the tree $\mathcal{T}$ satisfies $(\alpha, \frac{2\alpha \left\lceil \log_2(T)\right\rceil\bar{b}^2}{\sigma^2})-$RDP. Since $(s_1,\cdots, s_T)$ is obtained by the post-processing on the tree nodes, it also satisfies $(\alpha, \frac{2\alpha \left\lceil \log_2(T)\right\rceil\bar{b}^2}{\sigma^2})-$RDP. This implies $(s_1,\cdots, s_T)$ satisfies $(\frac{2\alpha \left\lceil \log_2(T)\right\rceil\bar{b}^2}{\sigma^2}+\frac{\ln(1/\delta)}{\alpha-1},\delta)$-DP with $\delta\in (0,1)$ and $\alpha>1$ and $(x_1,\cdots, x_T)$ satisfies $(\frac{2\alpha \left\lceil \log_2(T)\right\rceil\bar{b}^2}{\sigma^2}+\frac{\ln(1/\delta)}{\alpha-1},\delta)$-JDP by Billboard Lemma~\ref{lma:DP_JDP}. 
 Optimally choosing $\alpha=1+\sqrt{\frac{\ln(1/\delta)\sigma^2}{2\left\lceil \log_2(T)\right\rceil\bar{b}^2}}$, the composition $(x_1,\cdots, x_T)$ satisfies $(\frac{2\left\lceil \log_2(T)\right\rceil\bar{b}^2}{\sigma^2}+2\sqrt{\frac{2\left\lceil \log_2(T)\right\rceil\bar{b}^2}{\sigma^2}\ln(1/\delta)},\delta)$-JDP. Choosing $\sigma=\frac{\sqrt{2}\left\lceil \log_2(T)\right\rceil\bar{b}(\sqrt{\ln(1/\delta)}+\sqrt{\ln(1/\delta)+\epsilon})}{\epsilon}$, we can guarantee $(\epsilon,\delta)-$JDP.
\end{proof}

\subsection{Proof of Theorem~\ref{thm:reward_central}}\label{sec:proof_central}
Theorem~\ref{thm:reward_central} holds if the conditions in Theorem~\ref{thm:reward_meta} are satisfied, so we separately prove the two conditions.
\subsubsection{Proof of Condition (1)}
\begin{proof}
The clean remaining budget at round $t$ can be expressed as $B_t=B_1-\sum_{t=1}^Tb_t(x_t)$. Let $\mathcal{M}_t$ be the set of nodes that need to be aggregated by the $\mathrm{GetSum}(\mathcal{T},t)$ function in the tree-based aggregation scheme. We have the number of aggregated nodes is $|\mathcal{M}_t|\leq \lceil\log_2(t)\rceil$. Thus, the  privatized remaining budget is expressed as $\tilde{B}_{t}=B_1-s_t=B_{1}-\sum_{\tau=1}^{t-1}(b_{\tau}(x_{\tau}))-\sum_{i\in\mathcal{M}_t}n_{i}=B_t- \sum_{i\in\mathcal{M}_t}n_{i}$ where $n_i$ is the noise added to the tree node $i$. 

Let $v_t=\sum_{i\in\mathcal{M}_t}n_{i}$.Observing that the noise added to each node in $\mathcal{M}_t$ is an  independent Gaussian variable $\mathcal{N}(0,\sigma^2\bm{I})$, we have $v_t\sim \mathcal{N}(0,\log_2(t)\sigma^2\bm{I})$.  Given any $j\in [m]$, by Lemma~\ref{lma:Gaussian_concentration} and union bounds, given a resource $j\in [m]$, it holds $\forall t\in [T]$ with probability at least $1-\zeta $ that, 
\begin{equation}\label{eqn:sum_gaussian_bound_ftrl}
|v_{t,j}|\leq \sigma \sqrt{2\left\lceil\log_2(t)\right\rceil\ln\frac{2 T}{\zeta}},
\end{equation}
 By union bound, it holds  with probability at least $1-\zeta$ that  
 \begin{equation}\label{eqn:concentration_norm_inf}
   \| B_t-\tilde{B}_t\|_{\infty}=\| v_t\|_{\infty} \leq  \sigma \sqrt{2\left\lceil\log_2(t)\right\rceil\ln\frac{2 mT}{\zeta}}=B_{\mathrm{gap},t}.  
 \end{equation}

\end{proof}

\subsubsection{Proof of Condition (2)}

\begin{proof}
We first prove that if $\mu_t$ is updated by perturbed FTRL as $\mu_{t+1}=\arg\min_\mu (\sum_{\tau=1}^t g_{\tau})^\top\mu+ v_{t}^\top\mu+\frac{1}{\eta}h(\mu)$ where $\forall t\in[T], \|g_t\|\leq G$, then for any $\mu$, it holds for $w_t(\mu_t)=g_t^\top \mu_t$ that
\begin{equation}
\sum_{t=1}^T w_t(\mu_t)-w_t(\mu)\leq \eta (\frac{1}{2}G^2T+G\sum_{t=1}^T\|v_t\|) +\frac{1}{\eta}(h(\mu)-\min_{\beta>0}h(\beta)),
\end{equation}

Denote the non-perturbed FTRL update as $\mu'_{t+1}=\arg\min_\mu (\sum_{\tau=1}^t g_{\tau})^\top\mu+\frac{1}{\eta}h(\mu)$. Applying Lemma \ref{lma:lemma_7} with $\phi_1(\mu)=\eta(\sum_{\tau=1}^t g_{\tau})^\top\mu+h(\mu)$ and $\phi_2(\mu)=\eta (\sum_{\tau=1}^t g_{\tau})^\top\mu+h(\mu)+ \eta v_{t}^\top\mu$, we have
\begin{equation}\label{eqn:purturbed_deviation_proof}
\|\mu_{t+1}-\mu'_{t+1}\|\leq \eta \|v_t\|.
\end{equation}
Thus, we have 
\[
\begin{split}
&\sum_{t=1}^T w_t(\mu_t)-w_t(\mu)=\sum_{t=1}^T [g_t^\top \mu_t-g_t^\top \mu'_t +g_t^\top \mu'_t-g_t^\top \mu]\\
\leq &  \eta G\sum_{t=1}^T\|v_t\|+ \frac{\eta}{2} G^2 T+\frac{1}{\eta}(h(\beta)-\min_{\nu>0}h(\beta)),
\end{split}
\]
where the inequality holds by \eqref{eqn:purturbed_deviation_proof}, Cauchy-Schwartz inequality and Lemma \ref{lma:FTRL}. 

In Algorithm~\ref{alg:algorithm_ftrl}, we have $g_t=\rho-b_t(x_t)$, so  $G=\max\{\rho,\bar{b}-\rho\}$. Also, we have $v_t=\sum_{i\in\mathcal{M}_t}n_{i}$ in Algorithm~\ref{alg:algorithm_ftrl} where $\mathcal{M}_t$ is the set of nodes that need to be aggregated by the $\mathrm{GetSum}(\mathcal{T},t)$ function and $|\mathcal{M}_t|\leq \lceil\log_2(t)\rceil$. 

By \eqref{eqn:concentration_norm_inf}, we have with probability at least $1-\zeta$ that $\forall t\in [T]$,  $\|v_t\|_2\leq \sqrt{m}\|v_t\|_{\infty}\leq \sigma  \sqrt{2m\left\lceil\log_2(T)\right\rceil\ln(\frac{2mT}{\zeta})}$. 
Thus, it holds with probability at least $1-\zeta$ that 
\begin{equation}\label{eqn:ftrl_advproof_3}
\begin{split}
\sum_{t=1}^T w_t(\mu_t)-w_t(\mu)\leq \eta T ( \frac{G^2}{2}+ G\sigma  \sqrt{2m\left\lceil\ln(T)\right\rceil\ln(\frac{2mT}{\zeta})} )  +\frac{1}{\eta}(h(\mu)-\min_{\beta>0}h(\beta)).
\end{split}
\end{equation}

By choosing $\eta=\sqrt{\frac{h(\mu)-\min_{\beta>0}h(\beta)}{T(\frac{G^2}{2}+ G\sigma  \sqrt{2m\left\lceil\log_2(T)\right\rceil\ln(\frac{2mT}{\zeta})} )}}$, it holds with probability at least $1-\zeta, \zeta\in (0,1)$ that 
\begin{equation}
\begin{split}
\sum_{t=1}^T&w_t(\mu_t)-\sum_{t=1}^Tw_t(\mu)\\
&\leq \sqrt{T ( \frac{G^2}{2}+ G\sigma  \sqrt{2m\left\lceil\log_2(T)\right\rceil\ln(\frac{2mT}{\zeta})} )(h(\mu)-\min_{\beta>0}h(\beta))}
=\widetilde{\mathcal{O}}(\frac{\sqrt{T\sqrt{\log(1/\delta)}}}{\sqrt{\epsilon}}),
\end{split}
\end{equation}
where $G=\max\{\rho,\bar{b}-\rho\}$, and $\mu\in \{0, \frac{\bar{f}}{\alpha \rho_1}\mathbf{e}_1,\cdots, \frac{\bar{f}}{\alpha \rho_m}\mathbf{e}_m\}$ where $m$ is the number of constraints, $\mathbf{e}_j\in \{0,1\}^m$ with 1 at $j-$th entry and 0 at the other entries.

\end{proof}
By \eqref{eqn:concentration_norm_inf} in the proof of Condition~(1), to ensure \(B_{\mathrm{gap},t}=\sigma \sqrt{2\left\lceil\log_2(t)\right\rceil\ln\frac{2 mT}{\zeta}}\leq \Gamma(T,\zeta)\) and the dual convergence bound in Condition~(2), we can choose 
\begin{equation}
\begin{split}
\Gamma(T,\zeta)=&\sqrt{T ( \frac{G^2}{2}+ G\sigma  \sqrt{2m\left\lceil\log_2(T)\right\rceil\ln(\frac{2mT}{\zeta})} )(h(\mu)-\min_{\beta>0}h(\beta))}\\
&+\sigma \sqrt{2\left\lceil\log_2(T)\right\rceil\ln\frac{2 mT}{\zeta}},
\end{split}
\end{equation} 
where we set $\sigma=\frac{\sqrt{2}\bar{b}\sqrt{\lceil\log_2(T)\rceil}(\sqrt{\ln(1/\delta)}+\sqrt{\ln(1/\delta)+\epsilon})}{\epsilon}$ to preserve $(\epsilon,\delta)$-JDP. 

Moreover, by choosing $\zeta=\zeta_T=\frac{1}{T^{\vartheta}+1}$ given a constant $\vartheta\geq 1$, we have \[\Gamma(T,\zeta_T)=\mathcal{O}(\sqrt{T\ln(T)}\frac{\sqrt{\log(1/\delta)}}{\epsilon})=\widetilde{\mathcal{O}}(\frac{\sqrt{T\sqrt{\log(1/\delta)}}}{\sqrt{\epsilon}}).\]
Therefore, conditions in (2) of Theorem~\ref{thm:reward_meta} are satisfied. 

\section{Important Lemmas}\label{sec:importantlemma}
\subsection{Lemmas about privacy}
\begin{lemma}\label{Gaussian_Mechnism}
Gaussian Mechanism $\mathcal{A}(\mathcal{D}):= p(\mathcal{D})+\mathcal{N}(0,\sigma^2 \mathbf{I})$, where $\sigma=\Delta\sqrt{2\log(1.25/\delta)}/\epsilon$ with $\Delta:=\sup_{\mathcal{D}\sim \mathcal{D}'}\|p(\mathcal{D})-p(\mathcal{D}')\|_2$ as the global sensitivity of $p:\mathcal{D}\rightarrow \mathcal{Z}$,  is $(\epsilon,\delta)$-DP.
\end{lemma}

\begin{lemma}[Billboard Lemma \cite{private_matching_allocation_hsu2014private}]\label{lma:DP_JDP}
 Suppose $\mathcal{M}:\mathcal{Y}^T\rightarrow \mathcal{Z}$ is $(\epsilon, \delta)-$DP. Consider any set of functions $\pi_t:\mathcal{Y}_t\times \mathcal{Z}\rightarrow \mathcal{X}_t, t\in [T]$ where $\mathcal{Y}_t$ is input space of user $t$. Then the composition $(\pi_1(\mathcal{Y}_1, \mathcal{M}(\mathcal{Y}^T)), \cdots, \pi_t(\mathcal{Y}_t, \mathcal{M}(\mathcal{Y}^T)))$ is $(\epsilon, \delta)-$ joint differential private.
\end{lemma}
Lemma~\ref{lma:DP_JDP} shows that a JDP preserving algorithm can decide an action $x_t$ for user $t$ based on the true context $y_t$ and a privatized information $z$ from other users, using a non-privatized policy $\pi_t$.  Thus, JDP-preserving algorithms offer much more flexibility for reward optimization than DP.  The proposed algorithms in Section~\ref{sec:local_algrithm} preserve $(\epsilon, \delta)-$JDP.

\begin{lemma}[Relationship between LDP and JDP]\label{lma:LDP-JDP}
 If each local privacy preserving mechanism $\mathcal{M}_t: \mathcal{Y}_t\rightarrow \mathcal{Z}_t$ at user $t$ satisfies ($\epsilon, \delta$)-LDP, then an algorithm $\mathcal{A}:\mathcal{Y}^T\rightarrow \mathcal{X}^T$ which combines the same local processing and privacy preserving mechanisms $\{\mathcal{M}_t,t\in[T]\}$ at a trusted central node satisfies ($\epsilon, \delta$)-JDP. 
\end{lemma}
Lemma~\ref{lma:LDP-JDP} directly holds by Billboard Lemma in Lemma \ref{lma:DP_JDP} which establishes that a JDP preserving algorithm can select action for a user $t$ based on the corresponding context $y_t$ and the privatized information $z$ from other users. This fact is useful for offloading the computation burden from local users when a trusted central node is available.

\begin{lemma}[\cite{Renyi_DP_mironov2017renyi}]\label{Gaussian_Mechnism_RDP}
Gaussian Mechanism $\mathcal{M}(\mathcal{D}):= f(\mathcal{D})+\mathcal{N}(0,\sigma^2 \mathbf{I})$ is $(\alpha, \frac{\alpha \cdot \Delta_f^2}{2\sigma^2})$-RDP where $\Delta_f:=\sup_{\mathcal{D}\sim \mathcal{D}'}\|f(\mathcal{D})-f(\mathcal{D}')\|_2$ is the global sensitivity of $f:\mathcal{D}\rightarrow \mathcal{R}$.
\end{lemma}

\begin{lemma}[\cite{Renyi_DP_mironov2017renyi}]\label{lma:DP-RDP}
If $\mathcal{M}$ is $(\alpha, \epsilon)-$RDP, it is also $(\epsilon+\frac{\ln(1/\delta)}{\alpha-1}, \delta)-DP$ for any $\delta\in (0,1)$.
\end{lemma}

\begin{lemma}\label{lma:RDP_dependent_composition}[Proposition 1 in \cite{Renyi_DP_mironov2017renyi}]
Let $\mathcal{M}_1: \mathcal{D}\rightarrow \mathcal{R}_1$ be $(\alpha,\epsilon_1)-$RDP and $\mathcal{M}_2:\mathcal{D}\times \mathcal{R}_1\rightarrow \mathcal{R}_2$ be $(\alpha, \epsilon_2)-RDP$. Let $X:= \mathcal{M}_1(\mathcal{D})$ be the output of $\mathcal{M}_1$ and $Y:= \mathcal{M}_2(\mathcal{D},X)$ be the output of $\mathcal{M}_2$, then the composition $(X,Y)$ satisfies $(\alpha, (\epsilon_1+\epsilon_2)-$RDP. 
\end{lemma}

\begin{lemma}\label{lma:RDP_composition}
Let $\mathcal{M}_1: \mathcal{D}_1\rightarrow \mathcal{R}_1$ be $(\alpha,\epsilon_1)-$RDP and $\mathcal{M}_2:\mathcal{D}_2\times \mathcal{R}_1\rightarrow \mathcal{R}_2$ be $(\alpha, \epsilon_2)-RDP$. $\mathcal{D}_1$ and $\mathcal{D}_2$ are disjoint. Let $X:= \mathcal{M}_1(\mathcal{D}_1)$ be the output of $\mathcal{M}_1$ and $Y:= \mathcal{M}_2(\mathcal{D}_2,X)$ be the output of $\mathcal{M}_2$, then the composition $(X,Y)$ satisfies $(\alpha, \max \{\epsilon_1, \epsilon_2\})-$RDP. 
\end{lemma}
\begin{proof}
Let $P_1$ be the distribution of $X$, $P_2$ be the distribution of $Y$, and $P$ is the joint distribution of $(X,Y)$ given the dataset $\mathcal{D}=\{\mathcal{D}_1,\mathcal{D}_2\}$. Let $\mathcal{D}'$ be a neighboring dataset of $\mathcal{D}$, i.e. either $\mathcal{D}'=\{\mathcal{D}'_1,\mathcal{D}_2\}$ or $\mathcal{D}'=\{\mathcal{D}_1,\mathcal{D}'_2\}$. Given the neighboring dataset $\mathcal{D}'$, denote $P_1'$ be the distribution of $X'$, $P_2'$ be the distribution of $Y'$, and $P'$ is the joint distribution of $(X',Y')$. Then we have
\begin{equation}
\begin{split}
\exp&\left[ (\alpha-1)D_{\alpha}(P||P')\right]=\mathbb{E}_{P'}\left[(\frac{P(x,y)}{P'(x,y)})^{\alpha}\right]\\
&=\int_{\mathcal{R}_1\times \mathcal{R}_2} \left(P(x,y)\right)^{\alpha}\left(P'(x,y)\right)^{1-\alpha}\mathrm{d}x\mathrm{d}y\\
&=\int_{\mathcal{R}_1\times \mathcal{R}_2} \left(P_1(x)P_2(y\mid x)\right)^{\alpha}\left(P_1'(x)P_2'(y\mid x)\right)^{1-\alpha}\mathrm{d}x\mathrm{d}y\\
&=\int_{\mathcal{R}_1} P_1(x)^\alpha P_1'(x)^{1-\alpha}\int_{ \mathcal{R}_2}\left(P_2(y\mid x)\right)^{\alpha}\left(P_2'(y\mid x)\right)^{1-\alpha}\mathrm{d}y \mathrm{d}x
\end{split}
\end{equation}
If $\mathcal{D}'=\{\mathcal{D}_1,\mathcal{D}_2'\}$, then $X$ and $X'$ share the same distribution, i.e. $P_1=P_1'$. Thus, we have
\begin{equation}
\begin{split}
\exp&\left[ (\alpha-1)D_{\alpha}(P||P')\right]=\int_{\mathcal{R}_1} P_1(x)\int_{ \mathcal{R}_2}\left(P_2(y\mid x)\right)^{\alpha}\left(P_2'(y\mid x)\right)^{1-\alpha}\mathrm{d}y \mathrm{d}x\\
&\leq \int_{\mathcal{R}_1} P_1(x)\exp((\alpha-1)\epsilon_2) \mathrm{d}x=\exp((\alpha-1)\epsilon_2),
\end{split}
\end{equation}
where the inequality holds because $\mathcal{M}_2$ is $(\alpha, \epsilon_2)-$RDP.

If $\mathcal{D}'=\{\mathcal{D}_1',\mathcal{D}_2\}$, then given the same $X$, the conditional distributions of $Y$ and $Y'$ are the same, i.e. $P_2(Y\mid X)=P_2'(Y'\mid X)$. Thus, we have
\begin{equation}
\begin{split}
\exp&\left[ (\alpha-1)D_{\alpha}(P||P')\right]=\int_{\mathcal{R}_1} P_1(x)^\alpha P_1'(x)^{1-\alpha}\int_{ \mathcal{R}_2}P_2(y\mid x)\mathrm{d}y\mathrm{d}x \\
&\leq \int_{\mathcal{R}_1} P_1(x)^\alpha P_1'(x)^{1-\alpha} \mathrm{d}x=\exp((\alpha-1)\epsilon_1),
\end{split}
\end{equation}
where the inequality holds because $\mathcal{M}_1$ is $(\alpha, \epsilon_1)-$RDP. 
Therefore, we have $D_{\alpha}(P||P')\leq \max(\epsilon_1,\epsilon_2)$ which translates to $(\alpha,\max(\epsilon_1,\epsilon_2))-$RDP of the composition $(X,Y)$.
\end{proof}

\subsection{Lemmas about concentration}

\begin{lemma}[Sub-Gaussian Concentration]\label{lma:Gaussian_concentration}
Given a sub-Gaussian variable $X$ with $\mathbb{E}\left[ e^{\lambda(X-\mathbb{E}[X])}\right]\leq \exp(\frac{\lambda^2\sigma^2}{2})$,  it holds that
\[
\Pr(|X-\mathbb{E}[X]|\geq y)\leq 2\exp(\frac{-y^2}{2\sigma^2}).
\]
\end{lemma}

\begin{lemma}[Azuma-Hoeffding Inequality]
Suppose $\{X_k:k=0,1,2,3\cdots\}$ is a martingale or super-martingale and $|X_k=X_{k-1}|\leq c_k$, then for any $y>0$ and integer $N$, it holds that
\[
\Pr[|X_N-X_0|\geq y]\leq 2\exp\left(\frac{-y^2}{2\sum_{k=1}^Nc_k^2}\right).
\]
\end{lemma}

\begin{lemma}[Laurent-Massart Bound]\label{lma:LM_bound}
Let $Z_1,\cdots, Z_k$ be independent, standard Gaussian random variables, and the sum of their squares is $X=\sum_{i=1}^kZ_i^2$. It holds that
\begin{equation}
   \mathbb{E}[X]=k,\quad\; \Pr(X-k\geq 2\sqrt{kx}+2x)\leq e^{-x}.
\end{equation}
\end{lemma}

\subsection{Lemmas on Online Learning}
\begin{lemma}[Convergence of DMD (Proposition 5 in \cite{dual_mirror_descent_online_allocation_balseiro2020dual})]\label{lma:DMD}
If $\mu_t$ is updated as $\mu_{t+1}=\arg\min_{\mu\in\mathcal{U}} g_t^\top\mu+\frac{1}{\eta}V_h(\mu,\mu_t)$ where $\mathcal{U}$ is a convex set and $\forall t\in [T], \|g_t\|_2\leq G$, then for any $\mu:\mathcal{U}$, it holds for $w_t(\mu_t)=g_t^\top \mu_t$  that
\begin{equation}
\sum_{t=1}^T w_t(\mu_t)-w_t(\mu)\leq \frac{\eta}{2} G^2 T+\frac{1}{\eta}V_h(\mu,\mu_1).
\end{equation}
\end{lemma}

\begin{lemma}[Convergence of FTRL (Extended from Corollary 7.9 in \cite{online_learning_orabona2019modern})]\label{lma:FTRL}
If $\mu_t$ is updated as $\mu_{t+1}=\arg\min_{\mu\in\mathcal{U}} \sum_{t=1}^T g_t^\top\mu+\frac{1}{\eta}h(\mu)$ where $\mathcal{U}$ is a convex set, $h(\mu)$ is a 1-strongly convex function w.r.t. $\|\cdot\|_2$, and $\forall t\in [T], \|g_t\|_2\leq G$, then for any $\mu:\mathcal{U}$, it holds for $w_t(\mu_t)=g_t^\top \mu_t$  that
\begin{equation}
\sum_{t=1}^T w_t(\mu_t)-w_t(\mu)\leq \frac{\eta}{2} G^2 T+\frac{1}{\eta}(h(\mu)-\min_{\beta\in\mathcal{U}}h(\beta)).
\end{equation}
\end{lemma}

\begin{lemma}[Lemma 7 in \cite{survey_online_learning_mcmahan2017survey}]\label{lma:lemma_7}
   Let $\phi_1:\mathcal{C}\rightarrow \mathcal{R}$ be a convex function such that $x_1=\arg\min_{x\in\mathcal{C}} \phi_1(x)$ exists. Let $\psi$ be a convex function such that $\phi_2(x)=\phi_1(x)+\psi(x)$ is 1-strongly convex w.r.t. norm $\|\cdot\|$. Let $x_2=\arg\min_{x\in\mathcal{C}}\phi_2(x)$. Then, for any $g\in\partial \psi(x_1)$, we have $\|x_1-x_2\|\leq \|g\|_*$ where $\|\cdot\|_*$ is the dual norm of $\|\cdot\|$.
\end{lemma}
 
\section{Details of Empirical Studies  and Additional Numerical Results}\label{sec:experiments_appendix}
\subsection{Problem Statement}
We consider a request routing problem in battery-powered edge AI systems \cite{Multi-agent-router_yue2025masrouter,edge_AI_service_placement_lin2021optimizing}. The rapid growth of AI has driven their deployment on resource-constrained edge devices to enable low-latency, real-time applications such as autonomous systems, industrial automation, and augmented reality \cite{edge_survey_luo2021resource,AI_edge_computing_tuli2023ai}. In such systems, multiple AI models with different capabilities and energy consumption profiles are deployed on battery-powered edge servers. As user requests arrive sequentially, the system must decide in real time which model should serve each request, balancing inference quality while satisfying the energy constraint.

This setting naturally leads to an online resource allocation problem, where routing decisions must be made without knowledge of future requests while ensuring that the cumulative energy consumption satisfies a global budget constraint. The shared energy budget introduces a side channel: decisions for one request may reveal information about others through the remaining resource state, raising privacy concerns.

Resource allocation can be performed in either centralized or decentralized settings. In a \emph{centralized} setting, edge servers are managed by a single entity, and resource allocation is performed by a trusted centralized controller with access to raw contextual information. In this case, joint differential privacy (JDP) is required to protect user information while enabling efficient coordination. In contrast, in a \emph{decentralized} setting, edge servers may belong to different entities, and no centralized allocator exists. The servers can only share limited coordination information to satisfy global constraints, and local differential privacy (LDP) is required to protect user data. Our framework addresses both settings, enabling privacy-preserving and efficient request routing under global resource constraints.

\subsection{Experiment Setups}\label{sec:experimentsetups}

\paragraph{Reward and consumption models}
Concretely, the objective of the problem is to maximize the overall inference quality, subject to the long term energy resource constraint. 
This online allocation problem has a finite horizon of $T$ rounds.
At each round $t \in \{1,\dots,T\}$, a user arrives with the prompt features
$y_t$. The allocator must choose an action from the set
$X_t = \{0,1,\dots,M\}$ with each element representing one of $M$ AI models or a null action $0$ that represents skipping this request.
Selecting a model $x_t \in \{1,\dots,M\}$ at round $t$ yields a reward
\(f_t(x_t)\) which depends on the $acc(x_t,y_t) \in \{0,1\}$ accuracy of the user prompt $y_t$ on the selected model $x_t$ based on an accuracy predictor. The null action gives $f_t(0)=0$.
Each action incurs a scalar resource consumption $b_t(x_t)$ 
The null action consumes no resource for this request, i.e., $b_t(0)=0$.
All actions are drawn from a shared energy resource with total budget
$B = \rho T$, where $\rho>0$ is the per-round budget rate. The realized
actions $x_t \in X_t$ must satisfy
\(
\sum_{t=1}^{T} b_t(x_t) \le B .
\)

\paragraph{Dataset}
We use the RouterBench dataset \cite{Routerbench_hu2024routerbench}, a benchmark designed for studying prompt routing across heterogeneous LLMs. Routerbench consists of 36,496 real user prompts paired with per-model evaluation outcomes. 

To get the inference accuracy for incoming prompts, we use a MLP model that, given a prompt embedding and a target LLM, outputs the probability of correctness. Prompt embeddings are generated using the \texttt{all-MiniLM-L12-v2} model from the \texttt{SentenceTransformer} library \cite{reimers-2019-sentence-bert}, which provides compact semantic representations of text. 
The MLP predictive model has $2$ hidden layers,
each with width 100.
The energy consumption data is generated via an empirical estimation function inspired by \cite{wilkins2024offline}:
\(
b_t(j)
=
\alpha_{j,0}\,\tau_t^{\mathrm{in}}
+
\alpha_{j,1}\,\tau_j^{\mathrm{out}}
+
\alpha_{j,2}\,\tau_t^{\mathrm{in}} \tau_j^{\mathrm{out}},
\)
where $\tau_t^{\mathrm{in}}$ is the input token length of the prompt at round
$t$, $\tau_j^{\mathrm{out}}$ is a model-specific output token cap, and
$\alpha_{j,0},\alpha_{j,1},\alpha_{j,2}>0$ are model-dependent coefficients. 

The dataset contains 111,524 requests in total. We convert it into an episodic dataset which contains 440 episodes, each with 4000 rounds. 

\textbf{Baselines}. 
We compare \ouralgdmd and \ouralgftrl against the following baselines. 

\textbf{Non-private baselines}:\\
\textbullet\;\textsf{\greedy}: greedy selects
the constraint-feasible action that directly maximizes the reward $f_t(x)$.\\
\textbullet\;\textsf{\dmd}: the dual mirror descent algorithm \cite{dual_mirror_descent_online_allocation_balseiro2020dual} with $h(\mu)$ being the squared norm function.\\
\textbullet\;\textsf{\ftrl}: update the dual variable with follow the regularized leader \cite{online_learning_orabona2019modern} with $h(\mu)$ being the squared norm function.

\textbf{Private baselines}:\\ \textbullet\;\textsf{\oadp}: a centralized online allocation algorithm preserving standard $(\epsilon,\delta)-$DP in Definition~\ref{def:DP} by additionally applying Gaussian mechanism to reward and consumption functions of \ouralgftrl. \\
\textbullet\;\textsf{\ndmd}: Noisy DMD \cite{noisy_DMD_chen2024noisy} adapted to satisfy the constraint feasibility in the same way as \ouralg without applying the projection in \ref{eqn:DMD}. \\
\textbullet\;\textsf{\nftrl}: Directly applying noisy FTRL without applying tree aggregation (TA) \cite{tree_aggredation_chan2011private} for consumption aggregation. \\
\textbullet\;\textsf{\pridown}: Algorithm 2 in \cite{JDP_packing_huang2018near} adapted to satisfy the constraint feasibility in the same way as \ouralg.

\paragraph{Performance metrics} 
The average performance is evaluated by the average reward normalized by the average optimal reward over the sequences in the dataset, i.e. $\mathrm{AVG}=\frac{\mathbb{E}[F^{\mathcal{A}}_T(\bm{y})]}{\mathbb{E}[OPT(\bm{y})]}$. The adversarial performance is evaluated by the empirical competitive ratio, i.e. $\mathrm{CR}=\inf \frac{F^{\mathcal{A}}_T(\bm{y})}{OPT(\bm{y})}$.

\paragraph{Default experimental setting}
In default setting of our experiments, each round corresponds to routing a user prompt from the
RouterBench dataset to one of $M=11$ real LLMs. Each resource allocation episode has
$T = 4000$ rounds. Rewards are rescaled to lie in $[0,1]$ and the energy consumption per prompt is no greater than $0.01 \mathrm{kWh}$. The battery capacity is set as $10 \mathrm{kWh}$ by default.
 The dual variable is initialized as
$\mu_1 = 0.$ The dual step size is fixed to $\eta = 2000.$ The maximum dual norm bound in \ouralgdmd is set as $\mu_{\max}=400$. If not stated, the privacy parameters are set as
$\varepsilon = 10$, $\delta = 0.01$.

The evaluations are conducted on a labtop (Macbook Air). It takes about 5 minutes to execute $\ouralg$ for all the sequences in the dataset.

\subsection{Additional Numerical Results}
\label{sec:additionalresults}

\begin{table*}[!t]
\scriptsize
\vspace{-0.2cm}
\centering
\resizebox{\textwidth}{!}{
\begin{tabular}{c|c c c c |c c c c } 
\toprule
\multirow{2}{*}{\textbf{($\delta$, $\epsilon$)}} & \multicolumn{4}{c|}{CR} & \multicolumn{4}{c}{AVG}\\ 
    & \textbf{\ouralgdmd} & \textsf{\ndmd} & \textbf{\ouralgftrl} & \textsf{\nftrl} & \textbf{\ouralgdmd} & \textsf{\ndmd} & \textbf{\ouralgftrl}    &\textsf{\nftrl}  \\ 
\midrule
(0.001, 2)  & \textit{0.228}&0.206 & \textbf{0.657} & 0.171 & \textit{0.440}  &0.377 & \textbf{0.735} & 0.322 \\
(0.001, 4)  & \textit{0.419}&0.395 & \textbf{0.765} & 0.336 & \textit{0.624}  &0.610 & \textbf{0.821} & 0.514 \\
(0.001, 10)   & \textit{0.632} & 0.601 & \textbf{0.832}  & 0.567 &\textit{0.783} &  0.763 & \textbf{0.874} &0.705   \\
(0.0001, 2)  & \textit{0.194}&0.174 & \textbf{0.627} & 0.145 & \textit{0.401}  &0.347 & \textbf{0.701} & 0.285 \\
(0.0001, 4)  & \textit{0.374}&	0.357 & \textbf{0.752} & 0.307 & \textit{0.593}  &0.580 & \textbf{0.810} & 0.480 \\
(0.0001, 10)   & \textit{0.612} & 0.581 & \textbf{0.828}  & 0.533 &\textit{0.768} &  0.748 & \textbf{0.870} &0.687   \\

\bottomrule
\end{tabular}
}
\caption{\small{The expected (AVG) and adversarial (CR) performance of algorithms under different JDP or LDP levels. \ndmd and \ouralgdmd admit both JDP and LDP, and \nftrl and \ouralgftrl admit only JDP. Bold font indicates the best performance under JDP, and Italic font indicates the best performance under LDP. }}
\label{table:delta_epsilon_ablation}
\end{table*}

\begin{figure*}[t]
    \centering
   \begin{subfigure}[b]{0.3\linewidth}
        \centering
        \includegraphics[width=1\linewidth]{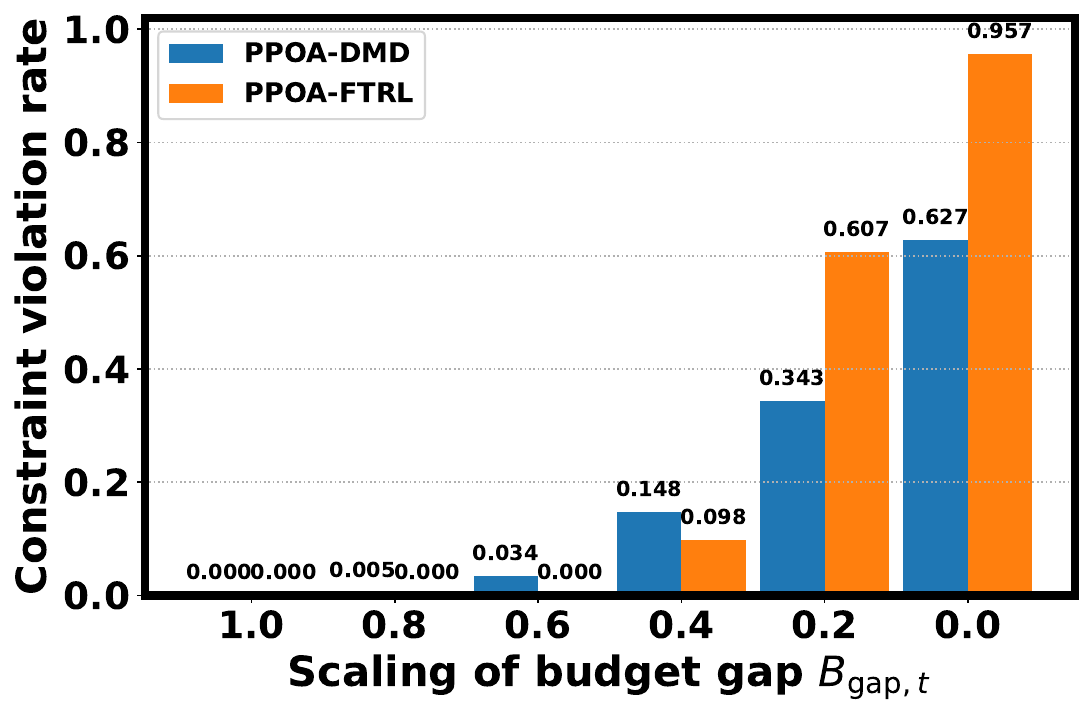}
        \caption{Resource constraint violation rate under different $B_{\mathrm{gap},t}$.}
        \label{fig:budgetgap-sweep}
    \end{subfigure}\hfill
    \begin{subfigure}[b]{0.3\linewidth}
        \centering
        \includegraphics[width=\linewidth]{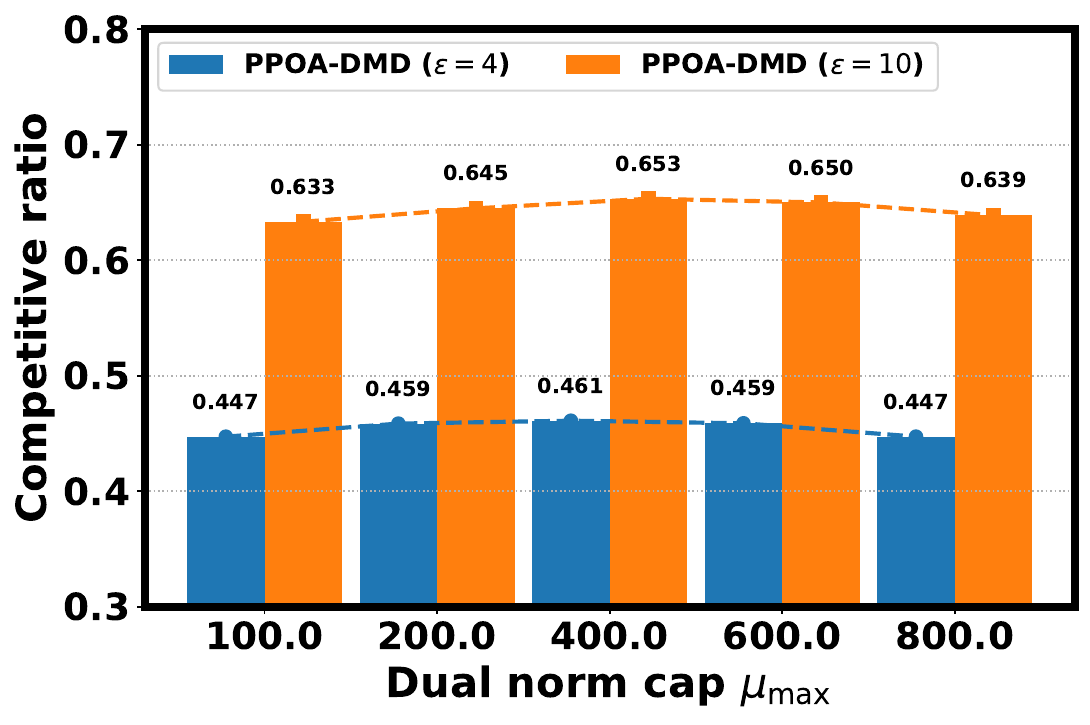}
        \caption{Competitive ratio of \ouralgdmd with $\mu_{\max}$.}
        \label{fig:mumax-sweep}
    \end{subfigure}\hfill
    \begin{subfigure}{0.31\textwidth}
        \centering
        \includegraphics[width=\linewidth]{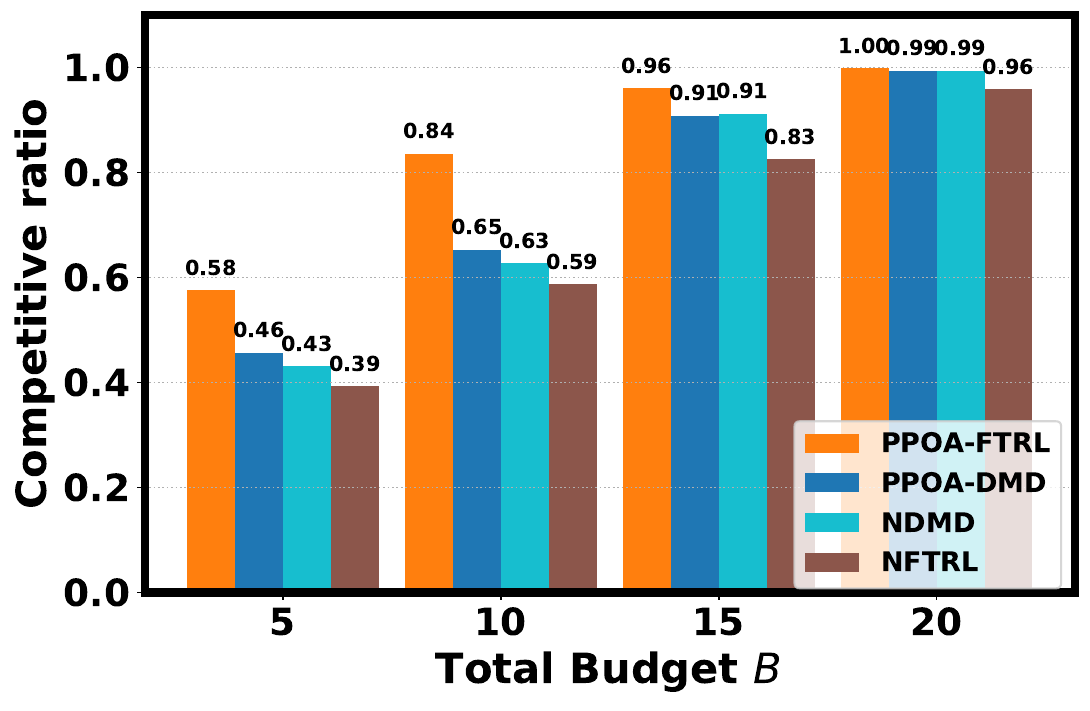}
        \caption{Competitive ratio under different budgets $B$.}
        \label{fig:budget-sweep}
    \end{subfigure}
    \caption{Ablation studies over different budget cap scaling, dual norm cap, and total budget. }
\end{figure*}

\paragraph{Ablation of Privacy Levels for Algorithms with $(\epsilon,\delta)-$JDP/LDP.}
In Table~\ref{table:delta_epsilon_ablation}, we further evaluate the average and adversarial performance of \ouralgdmd and \ndmd, which satisfy both JDP and LDP, as well as \ouralgftrl\ and \nftrl, which satisfy only JDP, under different combinations of $\delta$ and $\epsilon$. Since prior studies and practical applications \cite{apple_privacy,DL_differential_privacy_abadi2016deep,text_generation_dp_yue2023synthetic,noisy_DMD_chen2024noisy} commonly adopt $\delta$ that is close to zero and $\epsilon$ that lies between 1 and 10, we evaluate the algorithms on privacy levels with $\delta\in\{0.001,0.0001\}$ and $\epsilon\in\{2,4,10\}$. We observe that, for all algorithms, stronger privacy requirements (smaller $\delta$ or $\epsilon$) consistently lead to lower competitive ratios and average rewards, further illustrating the inherent trade-off between privacy and allocation performance. Moreover, across all combinations of $\delta$ and $\epsilon$, and for both adversarial and average performance metrics, \ouralgdmd\ consistently outperforms \ndmd\ under both JDP and LDP, while \ouralgftrl\ outperforms \nftrl\ under JDP. These results provide strong empirical evidence for the effectiveness of \ouralgdmd\ in mitigating the impact of noise through dual projection, as well as the advantage of \ouralgftrl\ in leveraging tree aggregation to achieve lower noise for the same privacy level.

Moreover, to assess the performance with very strong differential privacy, we evaluate our algorithms with privacy levels $\epsilon=\{0.1,0.2,0.5,1\}$. The results are provided in Table~\ref{table:delta_epsilon_ablation_strong_privacy}.  We can find that \ouralg still outperforms their counterparts with the same privacy notion under small enough privacy levels ($\epsilon=1$ and $\epsilon=0.5$). When we continue to increase the privacy guarantee to $\epsilon=0.2$ and $\epsilon=0.1$, the allocation performance guarantee becomes very weak. 

To examine performance under strong privacy requirements, we further evaluate the algorithms with $\delta=0.001$ and $\epsilon\in\{0.1,0.2,0.5,1\}$. As shown in Table~\ref{table:delta_epsilon_ablation_strong_privacy}, \ouralgdmd\ and \ouralgftrl\ outperform their respective baselines under the same privacy notions in both CR and AVG at $\epsilon=0.5$ and $\epsilon=1$. 
When $\epsilon$ decreases further to $0.2$ and $0.1$, all methods exhibit very low allocation performance. These results illustrate the substantial utility cost of imposing stringent privacy requirements at the evaluated horizon $T=4000$.
Nevertheless, a longer horizon can lead to greater utility under strong privacy because for any fixed privacy parameters, Theorem~\ref{thm:reward_meta} establishes that \ouralg\ achieves vanishing expected regret under i.i.d.\ arrivals and an asymptotic competitive ratio of at least $1/\alpha$ in
probability under adversarial arrivals.

\begin{table*}[!t]
\scriptsize
\vspace{-0.2cm}
\centering
\resizebox{\textwidth}{!}{
\begin{tabular}{c|c c c c |c c c c } 
\toprule
\multirow{2}{*}{\textbf{($\delta$, $\epsilon$)}} & \multicolumn{4}{c|}{CR} & \multicolumn{4}{c}{AVG}\\ 
    & \textbf{\ouralgdmd} & \textsf{\ndmd} & \textbf{\ouralgftrl} & \textsf{\nftrl} & \textbf{\ouralgdmd} & \textsf{\ndmd} & \textbf{\ouralgftrl}    &\textsf{\nftrl}  \\ 
\midrule
(0.001, 0.1)  & 0.0010&0.0010 & 0.0004 & 0.0006 & 0.0057  &0.0057 & 0.0003 & 0.0025 \\
(0.001, 0.2)  & 0.0044 &0.0045 & 0.0032 & 0.0030 & 0.0204  &0.0205 & 0.0007 & 0.0100 \\
(0.001, 0.5)   & 0.087 & 0.083 & \textbf{0.192}  & 0.053 &\textit{0.783} &  0.763 & \textbf{0.283} &0.018   \\
(0.001, 1)   & 0.088 & 0.084 & \textbf{0.470}  & 0.060 &\textit{0.241} &  0.221 & \textbf{0.567} &0.153 \\ 
\bottomrule
\end{tabular}
}
\caption{\small{The expected (AVG) and adversarial (CR) performance with strong privacy guarantee. \ndmd and \ouralgdmd admit both JDP and LDP, and \nftrl and \ouralgftrl admit only JDP. Bold font indicates the best performance under JDP, and Italic font indicates the best performance under LDP. Only scores larger than 0.1 are marked. }}
\label{table:delta_epsilon_ablation_strong_privacy}
\end{table*}

\paragraph{Effects of Reserved Budget Gap}
The reserved budget gap $B_{\mathrm{gap},t}$ in Line~4 of \ouralg\ (Algorithm~\ref{alg:algorithm_meta}) is critical for ensuring budget feasibility. In the default setting, we set $B_{\mathrm{gap},t}$ according to Theorem~\ref{thm:reward_local} for \ouralgdmd\ and Theorem~\ref{thm:reward_central} for \ouralgftrl, so that the resource constraints are satisfied. In Figure~\ref{fig:budgetgap-sweep}, we scale the reserved budget gap as $\vartheta \cdot B_{\mathrm{gap},t}$ across all rounds, where $\vartheta \in \{1.0, 0.8, 0.6, 0.4, 0.2, 0.0\}$. Here, $\vartheta=1$ corresponds to the default setting, while $\vartheta=0$ indicates no reserved budget gap. We observe that as $\vartheta$ decreases, the resource constraint violation rate increases rapidly, demonstrating the necessity of the budget gap design in \ouralg. Moreover, since \ouralgdmd\ injects more noise into the dual updates, it exhibits a higher violation rate than \ouralgftrl\ when the reserved budget gap is sufficiently large to be effective. In contrast, when the reserved budget gap is close to zero, \ouralgftrl\ shows a higher violation rate, likely because it is less conservative and tends to allocate more resources under this setting in the absence of a reserved safety margin.

\paragraph{Effects of $\mu_{\max}$ on \ouralgdmd}
$\mu_{\max}$ is a key parameter in the dual update of \ouralgdmd\ (Algorithm~\ref{alg:algorithm_local}), used to stabilize the dual learning process under noise injection for adversarial performance guarantees. According to Theorem~\ref{thm:reward_local}, $\mu_{\max}$ should be at least $\max_j \frac{\bar{f}}{\rho_j}$, which equals $100$ in our default experimental setting. In Figure~\ref{fig:mumax-sweep}, we evaluate the empirical competitive ratio of \ouralgdmd\ for $\mu_{\max} \in \{100, 200, 400, 600, 800\}$ under two privacy levels, $\epsilon=4$ and $\epsilon=10$. The results show that an overly small $\mu_{\max}$ degrades adversarial performance by overly restricting the dual domain and limiting the search for suitable dual variables. Conversely, an excessively large $\mu_{\max}$ also degrades performance, as it weakens the effect of regularizing noisy dual updates. Under this setting, $\mu_{\max}=400$ achieves the best performance for both $\epsilon=4$ and $\epsilon=10$.

\paragraph{Effect of Total Budgets}
The adversarial performance of \ouralg\ depends critically on the time-averaged budget $\rho = B/T$. We therefore evaluate the competitive ratio of \ouralg\ and other private baselines under different choices of $B$ with a fixed horizon $T=4000$, as shown in Figure~\ref{fig:budget-sweep}. All algorithms achieve zero constraint violation across all budget settings on the dataset, ensuring a fair comparison. As expected, all methods achieve better performance as the total budget increases. 

More importantly, under tighter budget constraints, \ouralgdmd\ exhibits larger improvements over \ndmd. This is because in resource-constrained regimes, the impact of injected noise on performance is amplified, making the effect of dual projection more critical. A similar trend is observed when comparing \ouralgftrl\ with \nftrl: under stricter budgets, the degradation caused by noise becomes more pronounced, and the superior information aggregation mechanism of \ouralgftrl\ yields greater gains. Overall, these results demonstrate the robustness and superiority of the proposed algorithms across varying budget regimes.

\section{Empirical Studies on Online Advertising}\label{sec:experiments_advertising}
\subsection{Problem Statement}
We evaluate private online allocation algorithms on a budget-constrained bidding problem over a horizon of \(T\) auctions in online advertising \cite{bidding_auction_online_advertising_aggarwal2024auto,liao2014ipinyou}. At round \(t\), an ad impression arrives with contextual information such as the user’s device and browser, geographic region, webpage domain, and ad-slot characteristics. An advertiser must decide a bid action \(x_t\in\mathcal{X}_t\) for an incoming ad opportunity at each round $t$. If the advertiser wins the auction for the ad opportunity, it earns the reward  \(f_t(x_t)\) related to the user click rate for this opportunity and pays the price \(b_t(x)\) for this opportunity. The null action \(x_t=0\) skips the auction and satisfies \(f_t(0)=b_t(0)=0\).  The total payment of the advertiser must not exceed a limited budget $B$ over a horizon $T$. 

Privacy matters because an impression’s context may reveal information about the user who generated it. That context can influence later bid actions observed by other users, which creates privacy leakage. We can protect from the privacy leakage by joint differential privacy. The resulting task is to earn as much reward as possible while respecting both the budget constraint and the privacy guarantee.

\subsection{Experiment Setups}

\paragraph{Dataset}
We evaluate the private online allocation algorithms on the iPinYou online advertising dataset \cite{liao2014ipinyou}. Each independent problem instance has horizon $T=4000$, and there are 600 instance sequences in the testing set. 

\paragraph{Reward and consumption models}
For each expression $t$, we fit a click probability model $\mathrm{pclick}_t$ based on the iPinYou dataset with impression features and corresponding click rate. We construct the synthetic market price sequence \(\{\mathrm{price}_t\}\) as a weighted combination of the scaled slot price, slot area, and hour, plus a constant baseline, multiplied by a Gaussian noise. 
For each nonzero action \(x\), the value attained from a click is set as \(\mathrm{vclick}_t=1\).  The winning probability of the non-zero biding action $x$ is modeled as \(\mathrm{win}_t(x)=\operatorname{sigmoid}((x\cdot\mathrm{vclick}_t \cdot\mathrm{pclick}_t -\mathrm{price}_t)/\tau)\), where \(\tau\) is half the standard deviation of the synthetic prices in the saved datasets. If the bid wins, the advertiser displays the ad and pays the modeled market price \(\{\mathrm{price}_t\}\). The expected reward given an action $x$ for impression $t$ is \(f_t(x)=\mathrm{win}_t(x)\cdot\mathrm{vclick}_t\cdot \mathrm{pclick}_t\) and expected expense is \(b_t(x)=\mathrm{win}_t(x)\cdot \mathrm{pclick}_t\). 

\paragraph{Baselines and performance metrics}
Baselines and performance metrics are the same as those of the main experiment in Section~\ref{sec:experimentsetups}.

\paragraph{Default setting}
In the default setting of our online advertising experiments, each round corresponds to an ad impression from the medium iPinYou dataset. The advertiser chooses among seven positive biding options and a no-bid action. We evaluate 600 episodes of \(T=4000\) rounds each. Rewards are rescaled to \([0,1]\), and the scaled expected spend per impression is at most \(0.1\). The total spending budget is \(B=100\) per episode.

In \ouralg algorithms, the dual variable is initialized at \(\mu_1=10\). The dual step size is set as \(\eta=2.5\) for PPOA-DMD and \(\eta=7.5\) for PPOA-FTRL using a validation dataset. The upper bound used in the PPOA-DMD dual projection is \(\mu_{\max}=5\). Unless otherwise stated, the privacy parameters are \(\varepsilon=10\) and \(\delta=0.001\).

\subsection{Numerical Results}
\begin{table*}[!t]
\scriptsize
\vspace{-0.2cm}
\centering
\resizebox{\textwidth}{!}{
\begin{tabular}{c|c c c| c c c c c c } 
\toprule
\multirow{2}{*}{\textbf{Algorithms}} & \multicolumn{3}{c|}{Non-private algorithms} & \multicolumn{5}{c}{Private algorithms}\\ 
    & \textsf{\greedy} & \textsf{\dmd} & \textsf{\ftrl} & \textsf{\oadp} & \textsf{\ndmd} & \textsf{\nftrl} & \pridown & \textbf{\ouralgdmd} & \textbf{\ouralgftrl}   \\ 
\midrule
AVG (10, 0.001)  & 0.691&0.986 & 0.986 & 0.641$\pm$0.0003& 0.599$\pm$0.001  &0.733$\pm$0.001 &0.769$\pm$0.0006 & 0.832$\pm$0.001 & \textbf{0.921}$\pm$0.0001 \\
CR (10, 0.001)  & 0.618 & 0.972 & 0.972  & 0.563$\pm$0.008 &  0.480$\pm$0.01 & 0.624$\pm$0.004 &  0.661$\pm$ 0.006&0.713$\pm$0.017 &\textbf{0.882}$\pm$0.004  \\
AVG (4, 0.001)  & 0.691&0.986 & 0.986 & 0.568$\pm$0.0002& 0.506$\pm$0.0015  &0.517$\pm$0.002 &0.516$\pm$0.0018 & 0.637$\pm$0.002 & \textbf{0.840}$\pm$0.0002 \\
CR (4, 0.001)  & 0.618 & 0.972 & 0.972  & 0.489$\pm$0.006 &  0.351$\pm$0.004 & 0.318$\pm$0.02 &  0.358$\pm$ 0.014&0.405$\pm$0.027 &\textbf{0.782}$\pm$0.007  \\
\bottomrule
\end{tabular}
}
\caption{\small{The expected (AVG) and adversarial (CR) performance of algorithms in the application of online bidding. Among the private algorithms, \oadp admits $(\epsilon,\delta)-$DP, \ndmd and \ouralgdmd admit $(\epsilon,\delta)-$JDP or $(\epsilon,\delta)-$LDP, \nftrl and \ouralgftrl admit $(\epsilon,\delta)-$JDP}}
\label{table:comparison_all_results_bid}
\end{table*}

We provide the main performance evaluation for \ouralg and baselines in Table~\ref{table:comparison_all_results_bid}.  Without privacy constraints, \dmd\ and \ftrl\ achieve the highest AVG and CR values.
Among methods supporting both JDP and LDP, \ouralgdmd\
consistently outperforms \ndmd\ across both privacy settings
and performance metrics, supporting the benefit of dual projection in~\eqref{eqn:DMD} for controlling the impact of privacy noise. When only JDP is required, \ouralgftrl\ achieves
the best performance among all private algorithms. Although all private methods
experience performance degradation as $\epsilon$ decreases, \ouralgftrl\ maintains its advantage in both metrics. These results are consistent with the benefit of its tree-aggregation mechanism for accurately aggregating historical resource consumption. 
The adversarial performance of \pridown with a multiplicative dual update drops a lot under stronger privacy preservation with less $\epsilon$ due to the lack of adversarial performance guarantee.  Moreover, despite enforcing standard DP, the centralized algorithm \oadp maintain a CR because it can perform centralized tree aggregation to get a good consumption estimation. 
The adversarial performance of \pridown, which uses a multiplicative dual update and lacks an adversarial performance guarantee, deteriorates substantially under stronger privacy requirements with smaller $\epsilon$. Meanwhile, despite enforcing standard DP, the centralized algorithm \oadp\ retains a CR of $0.489$ at
$\epsilon=4$, outperforming several private baselines, due to the benefit of centralized
tree aggregation for estimating cumulative resource consumption in \oadp.

\section{More Details about Applications and Additional Examples}\label{sec:more_application}

\subsection{Task Offloading for Battery-Powered Edge Computing}
A battery-powered edge node (mobile device, IoT gateway, or on-vehicle ECU)
serves a stream of computation requests from heterogeneous users.
At round $t$, request $y_t = \{X_t, f_t, b_t\}$ specifies an admissible
offloading action set $X_t$, a reward $f_t$ encoding latency / accuracy / QoS
satisfaction, and an on-device energy cost $b_t$. Following \cite{edge_computing_energy_constraints_jiang2022joint,edge_survey_luo2021resource,yang2018mobile}, an action $x_t \in X_t$ jointly determines
the local-versus-edge offloading ratio, the uplink transmission power, the computing-resource share, and thereby fixes both the latency
profile and energy consumption $b_t(x_t)$. Subject to a finite battery $B$, the platform must
maximize cumulative reward online, instantiating the online allocation problem
in Eq.~(1) of our paper. Coordination across rounds is carried by two
quantities: the residual-budget state $\tilde B_t$ and the dual price $\mu_t$,
which together summarize the entire request history into a compact signal that
governs the per-request offloading rule (Algorithm~\ref{alg:algorithm_meta}, Line~3).

The system admits both centralized and decentralized modes. In the \emph{centralized} mode, a trusted
edge allocator collects each arriving user's context, runs the primal--dual
update on its behalf, and returns the assigned action; users observe only their
own action and never see $\tilde B_t$ or $\mu_t$ directly. JDP preservation is required under this mode. In the
\emph{decentralized} mode, edge servers from different parties are collocated and share the same battery, and a trusted allocator does not exist: each server makes its
own offloading decision locally and instead broadcasts coordination summaries
(privatized budget consumption, dual updates) so that other servers can
maintain consistent estimates of $\tilde B_t$ and $\mu_t$ and respect the
shared energy budget. LDP preservation is required under this mode. 

\paragraph{Threat Model.}
We consider a malicious user who participates in the system as an ordinary
tenant. The attacker observes only the actions assigned to its \emph{own}
requests; it has no direct access to other users' contexts, identities, or
actions. Nevertheless, because the
per-request offloading rule is a function of the residual budget
and the dual price, the action returned for the attacker's own request reflects
the current level of resource pressure. By comparing the actions assigned to
its requests against what would normally be expected for those requests, the
attacker can detect whether the shared resource is being consumed faster than
usual or not: if the allocation is smaller than typical---for instance, more requests
are offloaded to cloud---then the residual budget is being depleted rapidly,
which suggests that one or more heavy consumers are currently being served.

This signal alone does not reveal much about user information by itself. The risk arises when the attacker combines it with auxiliary
information about a target user. Two representative scenarios illustrate this.
First, if the attacker knows when a target user typically arrives at the
system, it can submit its own requests at those moments and infer, from the
allocations it receives, whether the target's jobs in that window are heavy or
light, leaking task-type information about the user. Second, if the attacker
knows that a target user is characterized by distinctively heavy (or
distinctively light) and time-sensitive (or less time-sensitive) jobs, the attacker can probe the system continuously with its own requests and watch for sustained periods of unusually tight (or
unusually loose) allocations; the timing of those periods reveals the target's
arrival pattern. In both cases the attacker leverages a small amount of
external knowledge to get a much more informative inference about a specific user.

The same logic applies in the decentralized deployment, with the difference
that the coordination signal $\tilde B_t, \mu_t$ is broadcast directly and the
attacker no longer needs its own action as a proxy. This is the resource-constraint-based side channel: the shared energy
constraint nonlinearly couples decisions across rounds, so any quantity used to
enforce that constraint becomes a transcript of the request history. 

\subsection{Energy Management in Multi-Tenant Data Centers.}
A multi-tenant (colocation) data center co-locates servers owned and operated
by independent tenants under a shared power and energy infrastructure.
Crucially, the operator does not control tenant servers~\cite{islam2016market,dai2015energy,chen2015greening},
so the system is \emph{purely decentralized}: no trusted central allocator
exists, and tenants must coordinate among themselves to respect a shared
global energy budget. During peak hours the facility relies on a shared energy
storage system whose capacity over the peak window defines this global budget
$B$ for all tenants. At round $t$, the workload of an arriving tenant request
is captured by $y_t = \{X_t, f_t, b_t\}$, where $X_t$ is the set of admissible
task-level configurations (server count, CPU frequency, batch size, scheduling
option), $f_t$ is a performance reward (throughput, SLA satisfaction,
revenue), and $b_t$ is the induced energy draw. Tenants coordinate power reduction during peak hours through a
clearing price, and our setting treats the per-tenant configuration choice
$x_t \in X_t$ as the action of local resource allocation problem subject to the
finite global budget $B$. Each tenant decides $x_t$ locally and broadcasts
coordination summaries---privatized energy consumption and dual (energy price) updates---so
that the remaining tenants can maintain consistent estimates of the residual
budget $\tilde B_t$ and the dual price $\mu_t$. This decentralized operating mode is precisely the LDP regime of PPOA.

\paragraph{Threat Model.}
We consider a malicious tenant that participates as an ordinary occupant of
the data center and runs the same protocol as everyone else but has a goal of infering other tenants' information. Because the
system is decentralized, the attacker directly observes the broadcast
coordination signal---the disseminated $\tilde B_t$ and $\mu_t$, or
equivalently the privatized per-round energy consumption used to update them.
It has no access to other tenants' raw workloads, server states, or
configurations, but the coordination signal itself reflects the current level
of energy pressure across the facility. By comparing the trajectory of
$\tilde B_t$ and $\mu_t$ against what would normally be expected---the budget
draining faster than usual, or the dual price climbing earlier in the peak
window than typical---the attacker can detect when the shared energy budget is
being depleted unusually quickly, which suggests that one or more
heavy-consuming tenants are currently active.

This signal alone is coarse, but it becomes informative when combined with
auxiliary knowledge of a target tenant. First, if the attacker knows the time
windows in which a target tenant typically runs jobs, it can monitor the
coordination signal during those windows and infer whether the target's
workload is heavy or light, leaking workload-type information such as training
versus inference, batch versus interactive, or normal versus burst regimes.
Second, if the attacker knows that the target tenant is characterized by
distinctively heavy (or distinctively light) workloads, it can monitor the
coordination signal continuously and watch for sustained periods of unusually
fast (or unusually slow) budget depletion; the timing of those periods reveals
the target's activity pattern, which can in turn be correlated with business
hours, release cycles, or operational events.

\subsection{Budget-Aware Tool Calls in Agent Service.}
A tool-augmented agent service handles user queries by deciding, for each
incoming query, which external tools to invoke (search engines, web APIs,
retrieval databases) and how many tool-call rounds to spend. Recent work on
budget-aware agent scaling~\cite{liu2025budget} shows that effective performance
under bounded resources requires explicit awareness of a long-term tool-call
cost budget $B$, motivating an online allocation formulation. At round $t$,
the query is captured by $y_t = \{X_t, f_t, b_t\}$, where $X_t$ is the set of
admissible tool-call strategies (tool subset, number of rounds, reasoning
depth), $f_t$ is the answer-quality reward, and $b_t$ is the induced token cost or tool-call
cost consumption. This is a \emph{centralized} setting: a trusted agent-service
operator runs the primal--dual update for each user and returns the assigned
strategy, while users never see $\tilde B_t$ or $\mu_t$ directly. It therefore
corresponds exclusively to the JDP regime of PPOA.

\paragraph{Threat Model.}
A malicious user submits queries as an ordinary user and observes only the
tool-call actions assigned to its own queries---which tools were
invoked, how many calls were made, the depth of the reasoning trace, etc. Because
the tool-call action is a function of $\tilde B_t$ and $\mu_t$, deviations
from what is normally expected for those queries (such as fewer tool calls than usual or
shallower retrieval) indicate that the
shared cost budget is under pressure and that one or more heavy consumers are
currently active. Combined with auxiliary knowledge, this becomes informative:
if the attacker knows when a target user typically queries, it can probe in
those windows and infer whether the target is currently issuing heavy or light
queries; if the attacker knows the target is characterized by distinctively
heavy (or light) queries, sustained periods of unusually constrained actions reveal the target's activity pattern. Algorithm~\ref{alg:algorithm_meta}
closes this side channel by preserving JDP guarantees described in our paper.

\section{Notation Table}
\begingroup
\small
\renewcommand{\arraystretch}{1.12}
\setlength{\tabcolsep}{5pt}
\begin{longtable}{@{}p{0.2\textwidth}p{\dimexpr0.8\textwidth-2\tabcolsep\relax}@{}}
\caption{Main Notations.}
\label{tab:notation}\\
\toprule
\textbf{Notation} & \textbf{Description} \\
\midrule
\endfirsthead
\multicolumn{2}{l}{\tablename\ \thetable\ (continued)}\\
\toprule
\textbf{Notation} & \textbf{Description} \\
\midrule
\endhead
\midrule
\multicolumn{2}{r}{Continued on the next page}\\
\endfoot
\bottomrule
\endlastfoot

\multicolumn{2}{@{}l}{\textbf{Online allocation model}}\\*
$T$, $t\in[T]$ & Time horizon and round index; $[T]=\{1,\ldots,T\}$. \\
$m$, $j\in[m]$ & Number of resources and resource index. \\
$\mathcal X_t$, $x_t$ & Available action set and selected action at round $t$. \\
$f_t(x)$, $b_t(x)$ & Reward and resource-consumption vector for action $x$ at round $t$. \\
$\bar f$, $\bar b_j$, $\bar b$ & Uniform bounds satisfying $0\leq f_t(x)\leq\bar f$, $0\leq b_{t,j}(x)\leq\bar b_j$, and $\|b_t(x)\|_2\leq\bar b$. \\
$y_t$, $\bm y$ & User context $y_t=(\mathcal X_t,f_t,b_t)$ and context sequence $\bm y=(y_1,\ldots,y_T)$. \\
$\mathcal Y_t$, $\mathcal Y^T$ & Round-$t$ context space and sequence space $\mathcal Y^T=\prod_{t=1}^T\mathcal Y_t$. \\
$B$, $\rho$ & Total resource-budget vector and per-round budget vector, with $B=T\rho$. \\
$B_t$ & True remaining budget before round $t$: $B_t=B-\sum_{s=1}^{t-1}b_s(x_s)$. \\
$\hat\rho$ & Smallest per-round resource budget, $\hat\rho=\min_{j\in[m]}\rho_j>0$. \\
$\mathcal A$, $F_T^{\mathcal A}(\bm y)$ & Online algorithm and its total reward, $F_T^{\mathcal A}(\bm y)=\sum_{t=1}^T f_t(x_t)$. \\
$\mathrm{OPT}(\bm y)$ & Optimal reward achievable with full knowledge of $\bm y$. \\
$\alpha$ & $\alpha=\max_{j\in[m]}\frac{\bar b_j}{\rho_j}$; the corresponding competitive ratio is $1/\alpha$. \\
\addlinespace

\multicolumn{2}{@{}l}{\textbf{Privacy and budget estimation}}\\*
$\epsilon$, $\delta$ & Multiplicative privacy parameter and additive slack in the $(\epsilon,\delta)$ privacy inequality. \\
$\zeta$, $\zeta_T$ & Failure-probability parameter for the performance bounds and its horizon-dependent choice. \\
$\sigma$, $\sigma^2$ & Gaussian noise standard deviation and variance per coordinate. \\
$\tilde b_t(x_t)$ & Privatized consumption: $\tilde b_t(x_t)=b_t(x_t)+n_t$, with $n_t\sim\mathcal N(\bm0,\sigma^2 I_m)$. \\
$s_t$ & Privatized cumulative consumption through round $t$, released by tree aggregation. \\
$\tilde B_t$ & Privatized estimate of the remaining budget before round $t$. \\
$B_{\mathrm{gap},t}$ & Reserved budget margin bounding $\|\tilde B_t-B_t\|_\infty$ on the budget-estimation event. \\
$\mathcal M_B$, $\mathcal M_D$ & Privacy-preserving mechanisms for updating the remaining budget and dual variable. \\
\addlinespace

\multicolumn{2}{@{}l}{\textbf{Primal and dual updates}}\\*
$\mu_t$, $\mu$ & Nonnegative dual variable used at round $t$ and a dual comparator. \\
$\mathcal D$, $\mu_{\max}$ & Dual region and Euclidean norm bound used in the projected DMD update. \\
$\hat x_t$ & Pre-selected action before the budget check. \\
$w_t(\mu)$ & Linear dual loss, $w_t(\mu)=\mu^\top(\rho-b_t(x_t))$. \\
$f_t^*(\mu)$ & Maximized Lagrangian reward, $f_t^*(\mu)=\max_{x\in\mathcal X_t}\{f_t(x)-\mu^\top b_t(x)\}$. \\
$D(\mu\mid\bm y)$ & Offline dual objective, $D(\mu\mid\bm y)=\sum_{t=1}^T f_t^*(\mu)+T\mu^\top\rho$. \\
$\eta$, $h$ & Dual-update step size and strongly convex regularization function. \\
$V_h(\mu,\nu)$ & Bregman divergence, $V_h(\mu,\nu)=h(\mu)-h(\nu)-\nabla h(\nu)^\top(\mu-\nu)$. \\
$\mathbf e_j$ & $j$-th coordinate vector in $\mathbb R^m$. \\
\addlinespace

\multicolumn{2}{@{}l}{\textbf{Stochastic analysis}}\\*
$\mathcal P$, $\mathcal P_t$ & Common context distribution in the i.i.d.\ setting and marginal distribution of $y_t$ in the general non-stationary setting. \\
$\bar{\mathcal P}$ & Average marginal distribution, $\bar{\mathcal P}=\frac{1}{T}\sum_{t=1}^T\mathcal P_t$. \\
$\bar D(\mu\mid\mathcal P)$ & Expected dual objective per round, $\bar D(\mu\mid\mathcal P)=\mathbb E_{y\sim\mathcal P}[f_y^*(\mu)]+\mu^\top\rho$. \\
$v_t$, $\xi_t$ & Internal randomness used at round $t$ and history $\xi_t=\{(y_s,v_s):1\leq s\leq t\}$. \\
$\varsigma(\xi_t)$ & Sigma-algebra generated by the history through round $t$, including any initial randomness. \\
$\tau_A$ & First $s<T$ such that $\tilde B_{s+1,j}-B_{\mathrm{gap},s+1}\leq\bar b_j$ for some resource $j$. \\
$\mathcal E$ & Event on which both the budget-estimation bound and the dual-regret bound hold. \\
\addlinespace

\end{longtable}
\endgroup

\newpage
\section*{NeurIPS Paper Checklist}

The checklist is designed to encourage best practices for responsible machine learning research, addressing issues of reproducibility, transparency, research ethics, and societal impact. Do not remove the checklist: {\bf The papers not including the checklist will be desk rejected.} The checklist should follow the references and follow the (optional) supplemental material.  The checklist does NOT count towards the page
limit. 

Please read the checklist guidelines carefully for information on how to answer these questions. For each question in the checklist:
\begin{itemize}
    \item You should answer \answerYes{}, \answerNo{}, or \answerNA{}.
    \item \answerNA{} means either that the question is Not Applicable for that particular paper or the relevant information is Not Available.
    \item Please provide a short (1--2 sentence) justification right after your answer (even for \answerNA). 
\end{itemize}

{\bf The checklist answers are an integral part of your paper submission.} They are visible to the reviewers, area chairs, senior area chairs, and ethics reviewers. You will also be asked to include it (after eventual revisions) with the final version of your paper, and its final version will be published with the paper.

The reviewers of your paper will be asked to use the checklist as one of the factors in their evaluation. While \answerYes{} is generally preferable to \answerNo{}, it is perfectly acceptable to answer \answerNo{} provided a proper justification is given (e.g., error bars are not reported because it would be too computationally expensive'' or ``we were unable to find the license for the dataset we used''). In general, answering \answerNo{} or \answerNA{} is not grounds for rejection. While the questions are phrased in a binary way, we acknowledge that the true answer is often more nuanced, so please just use your best judgment and write a justification to elaborate. All supporting evidence can appear either in the main paper or the supplemental material, provided in appendix. If you answer \answerYes{} to a question, in the justification please point to the section(s) where related material for the question can be found.

IMPORTANT, please:
\begin{itemize}
    \item {\bf Delete this instruction block, but keep the section heading ``NeurIPS Paper Checklist"},
    \item  {\bf Keep the checklist subsection headings, questions/answers and guidelines below.}
    \item {\bf Do not modify the questions and only use the provided macros for your answers}.
\end{itemize} 

\begin{enumerate}

\item {\bf Claims}
    \item[] Question: Do the main claims made in the abstract and introduction accurately reflect the paper's contributions and scope?
    \item[] Answer: \answerYes{}
    \item[] Justification: The claims in the abstract and introduction are clearly justified by the theoretical analysis in Sections~\ref{sec:DP_metrics}, \ref{sec:meta_alg}, \ref{sec:local_algrithm}, and numerical results in Section~\ref{sec:experiments}.

\item {\bf Limitations}
    \item[] Question: Does the paper discuss the limitations of the work performed by the authors?
    \item[] Answer: \answerYes{}
    \item[] Justification: Limitations of the work are clearly discussed in Section~\ref{sec:conclusion}

\item {\bf Theory assumptions and proofs}
    \item[] Question: For each theoretical result, does the paper provide the full set of assumptions and a complete (and correct) proof?
    \item[] Answer: \answerYes{}
    \item[] Justification: The full set of assumptions are specified at the start of Section~\ref{sec:formulation}. The proofs for all the theoretical results are provided in Sections~\ref{sec:proofs_lower_bound},~\ref{sec:proof_meta},~\ref{sec:proof_ppoa_dmd}, ~\ref{sec:proof_PPOA_ftrl}, and \ref{sec:importantlemma}.

    \item {\bf Experimental result reproducibility}
    \item[] Question: Does the paper fully disclose all the information needed to reproduce the main experimental results of the paper to the extent that it affects the main claims and/or conclusions of the paper (regardless of whether the code and data are provided or not)?
    \item[] Answer: \answerYes{}
    \item[] Justification: the paper fully disclose all the information needed to reproduce the main experimental results of the paper to the extent that it affects the main claims and/or conclusions of the paper in  Section~\ref{sec:experiments_appendix}.

\item {\bf Open access to data and code}
    \item[] Question: Does the paper provide open access to the data and code, with sufficient instructions to faithfully reproduce the main experimental results, as described in supplemental material?
    \item[] Answer: \answerNo{}
    \item[] Justification: Codes for the numerical results will be opensourced once accepted.

\item {\bf Experimental setting/details}
    \item[] Question: Does the paper specify all the training and test details (e.g., data splits, hyperparameters, how they were chosen, type of optimizer) necessary to understand the results?
    \item[] Answer: \answerYes{}
    \item[] Justification: The dataset processing, hyperparameter choices and other setups are detailed in Section~\ref{sec:experimentsetups}.

\item {\bf Experiment statistical significance}
    \item[] Question: Does the paper report error bars suitably and correctly defined or other appropriate information about the statistical significance of the experiments?
    \item[] Answer:\answerYes{}
    \item[] Justification: The statistical variance of the algorithms is mainly from the injected privacy noise. The error bars for the main results are provided in Table~\ref{table:comparison_all_results}.

\item {\bf Experiments compute resources}
    \item[] Question: For each experiment, does the paper provide sufficient information on the computer resources (type of compute workers, memory, time of execution) needed to reproduce the experiments?
    \item[] Answer: \answerYes{}
    \item[] Justification: The comupter resource information is proved in Section~\ref{sec:experimentsetups}. 
    
\item {\bf Code of ethics}
    \item[] Question: Does the research conducted in the paper conform, in every respect, with the NeurIPS Code of Ethics \url{https://neurips.cc/public/EthicsGuidelines}?
    \item[] Answer: \answerYes{}
    \item[] Justification: The research conducted in the paper conform, in every respect, with the NeurIPS Code of Ethics.

\item {\bf Broader impacts}
    \item[] Question: Does the paper discuss both potential positive societal impacts and negative societal impacts of the work performed?
    \item[] Answer: \answerYes{}
    \item[] Justification: The paper discuss the societal impacts of the work performed in Section~\ref{sec:conclusion}.
    
\item {\bf Safeguards}
    \item[] Question: Does the paper describe safeguards that have been put in place for responsible release of data or models that have a high risk for misuse (e.g., pre-trained language models, image generators, or scraped datasets)?
    \item[] Answer: \answerNA{}
    \item[] Justification: The paper poses no such risks.

\item {\bf Licenses for existing assets}
    \item[] Question: Are the creators or original owners of assets (e.g., code, data, models), used in the paper, properly credited and are the license and terms of use explicitly mentioned and properly respected?
    \item[] Answer: \answerYes{}
    \item[] Justification: All original owners of assets including the datasets and models are properly credited and cited in Section~\ref{sec:experimentsetups}.

\item {\bf New assets}
    \item[] Question: Are new assets introduced in the paper well documented and is the documentation provided alongside the assets?
    \item[] Answer: \answerYes{}
    \item[] Justification: New assets introduced in the paper are well documented and the documentation is provided alongside the assets.

\item {\bf Crowdsourcing and research with human subjects}
    \item[] Question: For crowdsourcing experiments and research with human subjects, does the paper include the full text of instructions given to participants and screenshots, if applicable, as well as details about compensation (if any)? 
    \item[] Answer: \answerNA{}
    \item[] Justification: The paper does not involve crowdsourcing nor research with human subjects.

\item {\bf Institutional review board (IRB) approvals or equivalent for research with human subjects}
    \item[] Question: Does the paper describe potential risks incurred by study participants, whether such risks were disclosed to the subjects, and whether Institutional Review Board (IRB) approvals (or an equivalent approval/review based on the requirements of your country or institution) were obtained?
    \item[] Answer:\answerNA{}
    \item[] Justification: The paper does not involve crowdsourcing nor research with human subjects.

\item {\bf Declaration of LLM usage}
    \item[] Question: Does the paper describe the usage of LLMs if it is an important, original, or non-standard component of the core methods in this research? Note that if the LLM is used only for writing, editing, or formatting purposes and does \emph{not} impact the core methodology, scientific rigor, or originality of the research, declaration is not required.
    \item[] Answer: \answerNA{}
    \item[] Justification: This research does not involve LLMs as any important, original, or non-standard components

\end{enumerate}

\end{document}